\documentclass[prd,superscriptaddress,preprintnumbers,twocolumn,nofootinbib,showpacs]{revtex4-2}

\usepackage{amsfonts,amssymb,amsmath}
\usepackage{mathrsfs}
\usepackage{color}
\usepackage{graphicx}
\usepackage{dcolumn}
\usepackage{bm}
\usepackage{lipsum}
\usepackage{hyperref}

\begin{document}

\title{Primordial black holes and gravitational waves from \\ nonminimally coupled supergravity inflation}

\author{Shinsuke Kawai}
\email{kawai@skku.edu}
\affiliation{
	Department of Physics, 
	Sungkyunkwan University,
	Suwon 16419, Republic of Korea}
\author{Jinsu Kim}
\email{jinsu.kim@cern.ch}
\thanks{Corresponding Author}
\affiliation{
	School of Physics Science and Engineering, 
	Tongji University,
	Shanghai 200092, China}
\affiliation{
	Theoretical Physics Department,
	CERN,
	1211 Geneva 23, Switzerland} 
\date{\today} 
\preprint{CERN-TH-2022-157}

\begin{abstract}
We study the formation of primordial black holes and the generation of gravitational waves in a class of cosmological models that are direct supersymmetric analogs of the observationally favored nonminimally coupled Higgs inflation model. It is known that this type of model naturally includes multiple scalar fields which may be regarded as the inflaton. For the sake of simplicity we focus on the case where the inflaton field space is two dimensional. We analyze the multifield dynamics and find the region of parameters that gives copious production of primordial black holes that may comprise a significant part of the present dark matter abundance. We also compute the spectrum of the gravitational waves and discuss their detectability by means of future ground-based and space-borne gravitational wave observatories.
\end{abstract}

\maketitle

\section{Introduction}
In the radiation dominant era of the early Universe, the horizon mass in a Hubble volume may collapse to form a black hole if the density contrast is large enough.
Such primordial black holes (PBHs) can exist in a much wider mass spectrum than the black holes resulting from the standard stellar evolution, and are a subject of active investigation since they were first postulated in the 1960s \cite{Zeldovich:1967lct,Hawking:1971ei,Carr:1974nx,Polnarev:1985btg}.
After the direct detection of gravitational waves by the LIGO and Virgo collaborations \cite{LIGOScientific:2016aoc,LIGOScientific:2016sjg,LIGOScientific:2017ycc}, PBHs received renewed interest as several groups suggested that the source of the detected gravitational waves could be binary black holes of primordial origin; see, {\it e.g.}, \cite{Carr:2020xqk,Khlopov:2008qy,Sasaki:2018dmp,Carr:2020gox,Green:2020jor,Villanueva-Domingo:2021spv} for a review.
PBHs interact only through gravitation and are an excellent candidate of dark matter. 
The abundance of PBHs is constrained by observation, {\it e.g.}, microlensing, as well as by theoretical consistency, {\it e.g.}, successful big bang nucleosynthesis.
While it is now well known that the PBHs of LIGO-Virgo event mass scale $\sim 30 M_\odot$ cannot account for the total dark matter abundance of the present Universe (see, {\it e.g.}, \cite{Sasaki:2016jop}), there remains a mass window \cite{Montero-Camacho:2019jte,Carr:2020gox} at the asteroid scale $3.5\times 10^{-17}-4\times 10^{-12}\,M_{\odot}$ in which the whole or a significant part of today's dark matter abundance may be attributed to PBHs. 

Production of PBHs requires a large density contrast, much larger than that at the horizon exit scale of the cosmic microwave background (CMB).
An important recent theoretical development is that if such large density contrast is present, gravitational waves are generated at the second order in perturbation theory \cite{Matarrese:1997ay,Mollerach:2003nq,Ananda:2006af,Baumann:2007zm}, which may be large enough to be detected by future gravitational wave observatories; see, {\it e.g.}, \cite{Sasaki:2018dmp,Gong:2019mui,Domenech:2019quo,Domenech:2020kqm,Domenech:2021ztg,Yuan:2021qgz} for a review. 
If such gravitational waves are to be detected in the future, that would certainly mark a new era of cosmology, but even the null result would also give important constraints on the physics of the early Universe.

The power spectrum of the density perturbation needed for the production of PBHs is estimated to be seven orders of magnitude larger than that at the CMB scale, and thus it is a major theoretical challenge to build a model that naturally realizes the necessary enhancement of the power spectrum at small scales.
Single field inflation with an inflection point is known to have the desired feature, and thus far many models for enhancement of the scalar power spectrum have been proposed;
in some models, the enhancement is strong enough and abundant PBHs are produced, so that the PBHs may be considered as the dark matter of the Universe. 
One example is the single field model involving nonminimal Gauss-Bonnet coupling \cite{Kawai:2021edk}, in which the inflection-point like structure is maintained by the balance between the potential term and the Gauss-Bonnet coupling term \cite{Kawai:2021bye}.

High energy theories like string theory and supergravity naturally involve multiple scalar fields, and thus it seems fruitful to investigate possible production mechanisms of PBHs using the rich structure of the multifield dynamics. 
Such studies, in turn, will be able to constrain the UV physics if PBHs are to be found in the future within predicted mass ranges. 
There are many studies on PBHs and/or the associated gravitational waves in multifield inflation models, including
\cite{Silk:1986vc,Yokoyama:1995ex,Randall:1995dj,Garcia-Bellido:1996mdl,Kawasaki:1997ju,Kawasaki:2012wr,Kohri:2012yw,Clesse:2015wea,Ando:2017veq,Pi:2017gih,Cheong:2019vzl,Ketov:2019mfc,Ashoorioon:2020hln,Gundhi:2020kzm,Gundhi:2020zvb,Palma:2020ejf,Braglia:2020taf,Braglia:2020eai,Aldabergenov:2020bpt,Spanos:2021hpk,Ketov:2021fww,Iacconi:2021ltm,Hooshangi:2022lao,Ashoorioon:2022raz,Boutivas:2022qtl,Cheong:2022gfc,Geller:2022nkr,Bhattacharya:2022fze}. 
Of these, the hybrid inflation type models are of particular interest, since many phenomenological models of particle physics indeed have such structure. 
A well-known drawback of the hybrid inflation model is that its original version predicts an observationally disfavored blue scalar spectrum $n_s\gtrsim 1$, and viable scenarios of hybrid inflation typically involve some correction terms, rendering the scenarios less predictive.

Below, we discuss production of PBHs and generation of gravitational waves in a supergravity-based scenario of cosmic inflation that seems to have eluded attention, but still has interesting features.
The model involves multiple fields, but in one limit it approaches a single field model in which the effective potential becomes identical to that of the nonminimally coupled Higgs inflation model \cite{Cervantes-Cota:1995ehs,Bezrukov:2007ep}, which predicts observationally favored scalar spectral index $n_s\sim 0.97$ and small tensor-to-scalar ratio $r\ll 1$.
Adjusting the parameters originating from the K\"{a}hler potential, the model resembles the hybrid inflation model; nevertheless the CMB spectrum still remains observationally favored.
Like in other known scenarios of inflationary cosmology that generate PBHs, some degree of fine tuning is necessary in our model, and we show that sufficient enhancement of the scalar power is achieved by tuning some noncanonical (higher order) terms in the K\"{a}hler potential.
We show this by carrying out detailed numerical study on two sets of benchmark parameter values.

We start in the next section with the description of the inflationary model. 
In Section \ref{sec:infdynamics} we show the dynamics of inflation, and in Section \ref{sec:curvPS} we comment on the behavior of the curvature power spectrum.
Section \ref{sec:PBHsGWs} shows our analysis of the PBH production and the gravitational wave spectrum. 
We conclude in Section \ref{sec:conc} with brief comments.

\section{Nonminimally coupled supergravity inflation}
\label{sec:model}
Let us start with the supergravity Lagrangian
\begin{align}
\mathcal{L} \supset \int d^4\,\theta
\, \phi^\dagger\phi\, K
+\Big\{\int d^2\theta\,\phi^3\, W+\text{h.c.}\Big\}
\,,
\end{align}
where $\phi$ is the conformal compensator, $W$ is the superpotential, and $K$ is the K\"{a}hler potential in the superconformal framework.
For our model it is essential that the superpotential includes a term
\begin{align}\label{eqn:W}
W\supset yS\overline{X}X\,,
\end{align}
where $(X, \overline X)$ are a vectorlike pair of superfields and $S$ is a singlet or an adjoint superfield under some gauge symmetry.
While this structure is absent in the MSSM, it is commonly found in supergravity embedding of particle theories beyond the Standard Model
\cite{Einhorn:2009bh,Ferrara:2010in,Ferrara:2010yw,Arai:2011nq,Kawai:2015ryj,Pallis:2011gr,Arai:2011aa,Arai:2012em,Kawai:2014doa,Kawai:2014gqa,Kawai:2021gap,Arai:2013vaa,Leontaris:2016jty,Okada:2017rbf,Kawai:2021tzc,Kawai:2021hvs}.
We allow the K\"{a}hler potential in the superconformal framework\footnote{
The K\"{a}hler potential in the Einstein frame is $\mathcal{K}=-3M_{\rm P}^2\ln\Phi$.
We choose the mass scale of the supergravity to be the reduced Planck scale
$M_{\rm P}=2.44\times 10^{18}$ GeV.
} to be slightly noncanonical, $K=-3M_{\rm P}^2\Phi$ with
\begin{align}
\Phi &= 1-\frac{1}{3M_{\rm P}^2}\left(|S|^2+|\overline X|^2+|X|^2\right)
+\frac{\gamma}{2M_{\rm P}^2}\left(\overline X X+\text{h.c.}\right)
\nonumber\\
&\quad
+\frac{\sqrt 2\kappa_3}{3M_{\rm P}^3}\left(S^2S^*+\text{h.c.}\right)
+\frac{\kappa_4}{3M_{\rm P}^4}|S|^4
+\frac{\kappa_6}{3M_{\rm P}^6}|S|^6\,.
\end{align}
The third term proportional to $\gamma$ gives rise to the nonminimal coupling, and the real constants $\kappa_3$, $\kappa_4$, and $\kappa_6$ represent the parameter freedom of the model that are adjustable as long as they are not ridiculously large.

Let us parametrize the flat direction $h$ and the singlet direction $s$ along the scalar components of the multiplets 
\begin{align}
X=\frac{h}{2},\quad
\overline X=\frac{h}{2},\quad
S=\frac{s}{\sqrt{2}}\,.
\end{align}
In many examples of concrete inflationary models, the inflaton trajectories are known to be stable along the real directions of $s$ and $h$, and we thus restrict $s$ and $h$ to take real values below.
Then the superpotential and the K\"{a}hler potential read, setting $M_{\rm P}=1$ henceforth,
\begin{align}
W&=\frac{y}{4\sqrt 2} sh^2\,,\\
\label{eqn:Phi}
\Phi &= 1
-\frac{1}{6} s^2
+\xi h^2
+\frac{\kappa_3}{3}s^3
+\frac{\kappa_4}{12}s^4
+\frac{\kappa_6}{24}s^6\,,
\end{align}
where
\begin{align}
\xi=\frac{\gamma}{4}-\frac{1}{6}\,.
\end{align}
Of course, other terms, such as the ones in the MSSM, may be present in the theory but these will be neglected below, with assumption that the initial vacuum expectation values of the scalar components of the multiplets other than $s$ and $h$ are sufficiently small.

The Lagrangian in the Einstein frame is obtained by the Weyl transformation of the metric. 
It reads
\begin{align}
\mathcal{L}_{\rm E}=\sqrt{-g}\left[
\frac{1}{2} R
-\frac{1}{2} G_{ab}g^{\mu\nu}
\partial_\mu\varphi^a\partial_\nu\varphi^b
-V_{\rm E}
\right]\,,
\end{align}
where $\varphi^a=(h,s)$ and the components of the field space metric are
\begin{align}\label{eqn:Ghh}
G_{hh}&=\frac{6\xi^2h^2+\Phi}{\Phi^2}\,,\\
\label{eqn:Ghs}
G_{hs}&=-\frac{\xi hs}{\Phi^2}\left(1-3\kappa_3 s-\kappa_4 s^2-\frac 34\kappa_6 s^4\right)\,,\\
\label{eqn:Gss}
G_{ss}&=\frac{1}{\Phi^2}\Big[
\frac{4 s^4 \left(2 \kappa_3^2+\kappa_4\right)
+\kappa_6 s^6 \left(8-8\kappa_3s-\kappa_4 s^2\right)}{48}
\nonumber\\
&\quad+\left(1+\xi h^2\right) \left(1-4 \kappa_3 s-2 \kappa_4
s^2
-\frac{9}{4} \kappa_6 s^4\right)
\Big]\,.
\end{align}
The two dimensional field space $(s,h)$ is curved.
The scalar potential in the Einstein frame is
\begin{align}\label{eqn:EFpot}
V_{\rm E}
=\frac{V_{\rm J}}{\Phi^2}\,,
\end{align}
where
\begin{align}
V_{\rm J}=\frac{y^2h^2s^2}{4}
+\frac{y^2 h^4}{4d(s)}
+\frac{y^2 h^4 s^2[
 	\frac{3}{2}\gamma
 	-f(s)
 	]^2}{24-6\gamma h^2+9\gamma^2 h^2 + g(s)}
\,,
\end{align}
with
\begin{align}
d(s) &= 4(1-4\kappa_3 s - 2\kappa_4 s^2)-9\kappa_6 s^4
\,,\\
g(s) &= \frac{8(2\kappa_3^2+\kappa_4)s^4+16\kappa_6 s^6 - 16\kappa_3\kappa_6 s^7 - 2\kappa_4 \kappa_6 s^8}{d(s)}
\,,\\
f(s) &= \frac{2(\kappa_3+\kappa_4 s)s+3\kappa_6 s^4}{d(s)}
\,.
\end{align}

Formalisms for multifield inflation have been developed by many authors, including \cite{Peterson:2010np,Sasaki:1995aw,Nakamura:1996da,Gordon:2000hv,GrootNibbelink:2001qt,Gong:2011uw,Kaiser:2012ak,Schutz:2013fua}. Here we simply list a few key equations; see, {\it e.g.}, Appendix A of \cite{Kawai:2015ryj} for more explanation.
The background evolution is described by the equations of motion for the multifields,
\begin{align}\label{eqn:KGeq}
\frac{D\dot\varphi^a}{dt}+3H\dot\varphi^a+G^{ab}\nabla_bV(\varphi^c)=0\,,
\end{align}
together with the Friedmann equation
\begin{align}\label{eqn:Friedmann}
3H^2=\frac{1}{2} G_{ab}\dot\varphi^a\dot\varphi^b+V(\varphi^c)\,.
\end{align}
The overdot is the derivative with respect to the cosmic time, $D$ stands for covariantized time derivative, and $\nabla$ stands for the covariantized field space derivative.

On a given background, the dynamics of the perturbations is found by solving the equations
\begin{align}
&\ddot{Q}_\sigma + 3H\dot{Q}_\sigma + \left[
\frac{k^2}{a^2} + \mathcal{M}_{\sigma\sigma} - \omega^2
-\frac{1}{a^3}\frac{d}{dt}\left(
\frac{a^3\dot{\sigma}^2}{H}
\right)\right]Q_\sigma
\nonumber\\
&=
2\frac{d}{dt}\left(\omega Q_s\right)
-2\left(
\frac{V_{,\sigma}}{\dot{\sigma}} + \frac{\dot{H}}{H}
\right)\omega Q_s
\,,\\
&\ddot{Q}_s + 3H\dot{Q}_s + \left[
\frac{k^2}{a^2} + \mathcal{M}_{ss} + 3\omega^2
\right]Q_s 
\nonumber\\
&=
4\frac{\omega}{\dot{\sigma}}\frac{\dot{H}}{H}\left[
\frac{d}{dt}\left(\frac{H}{\dot{\sigma}}Q_\sigma\right)
-\frac{2H}{\dot{\sigma}}\omega Q_s
\right]
\,,\label{eqn:Qspert}
\end{align}
where $Q_\sigma$ and $Q_s$ are respectively the adiabatic and entropic perturbations, and
\begin{align}
\dot{\sigma} &\equiv \sqrt{G_{ab}\dot{\varphi}^a\dot{\varphi}^b}
\,,\\
\hat{\sigma}^a &\equiv \frac{\dot{\varphi}^a}{\dot{\sigma}}
\,,\\
\omega^a &\equiv
\frac{d\hat{\sigma}^a}{dt} + \Gamma^a_{bc}\dot{\varphi}^b\hat{\sigma}^c
\,,\\
\hat{s}^a &\equiv 
\frac{\omega^a}{|\omega^a|}
\,,\\
\mathcal{M}_{\sigma\sigma} &\equiv 
\hat{\sigma}_a\hat{\sigma}^b\mathcal{M}^a_b
\,,\\
\mathcal{M}_{ss} &\equiv
\hat{s}_a\hat{s}^b\mathcal{M}^a_b
\,,\\
\mathcal{M}^a_b &\equiv
G^{ac}\nabla_b \nabla_c V
-\mathcal{R}^a{}_{cdb}\dot{\varphi}^c\dot{\varphi}^d
\,,\\
V_{,\sigma} &\equiv \hat{\sigma}^a\nabla_aV
\,.
\end{align}
Here $\mathcal{R}^a{}_{cdb}$ is the field space Riemann tensor.
The curvature and isocurvature perturbations are then defined as follows:
\begin{align}
\mathcal{R} = \frac{H}{\dot{\sigma}}Q_\sigma
\,,\qquad
\mathcal{S} = \frac{H}{\dot{\sigma}}Q_s\,.
\end{align}

\section{Inflaton trajectories}
\label{sec:infdynamics}
We focus on the ``cubic K\"{a}hler potential model'' where $\kappa_6$ is set to be zero and the ``sextic K\"{a}hler potential model'' where $\kappa_3 = 0$.
Furthermore we assume that the inflationary trajectory starts in the large-$h$ limit with a small $s$-field value. In other words we study effects of the $s$ field on the otherwise standard single field nonminimal inflation model.\footnote{
Taking $s=0$, we recover the standard nonminimally coupled Higgs inflation which is known to give observationally favored prediction of the CMB spectrum.
}

For both cubic and sextic cases we initially have eight free parameters; four potential parameters $(\kappa_3 \text{ or }\kappa_6, \kappa_4, \xi, y)$, two initial field values $(h_{\rm ini}, s_{\rm ini})$, and two initial velocities $(\dot{h}_{\rm ini}, \dot{s}_{\rm ini})$. For the initial velocities, we use the slow roll background equations of motion which is a good approximation in the large-$h$ limit.
The parameter $y$ enters in the potential as an overall factor. We thus use $y$ to match the magnitude of the curvature power spectrum at the pivot scale, namely $\mathcal{P}_\zeta(k_*) \approx 2 \times 10^{-9}$. 
The current observational bound on the isocurvature perturbations is safely satisfied in our model.
Finally, the initial $s$-field value is chosen to be at the potential minimum at $h=h_{\rm ini}$, and $h_{\rm ini}$ is set to be 0.12 which is enough to achieve 60 $e$-folds. We are thus left with three parameters $(\kappa_3\text{ or }\kappa_6, \kappa_4, \xi)$.
For simplicity, we assume that after inflation the Universe undergoes immediate transition to radiation domination\footnote{
The effects of the delay of thermalization due to non-instantaneous reheating would only change the number of $e$-folds from our choice of 60.
Deviation of the $e$-folding number would slightly change the value of the $y$ parameter and hence the spectral index $n_s$ and the tensor-to-scalar ratio $r$. However, for the class of inflationary models we consider, the change is known to be small and safely within current observational bounds of the latest Planck \cite{Planck:2018jri} and BICEP/Keck \cite{BICEP:2021xfz} observations. In single-field models, this is discussed for example in Ref.~\cite{Kawai:2021hvs}.
}.

\subsection{Cubic case}
\begin{figure*}[ht!]
\centering
\includegraphics[width=0.3\textwidth]{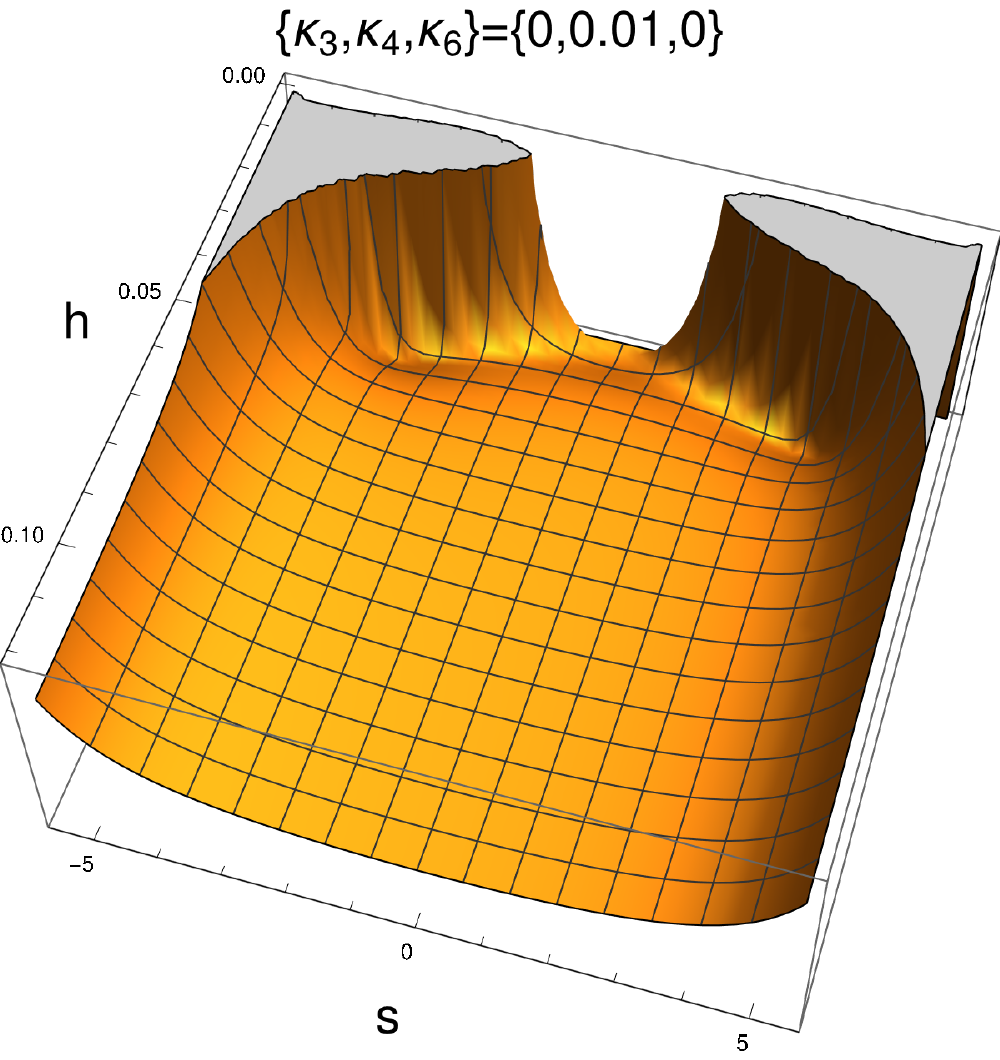} 
\includegraphics[width=0.3\textwidth]{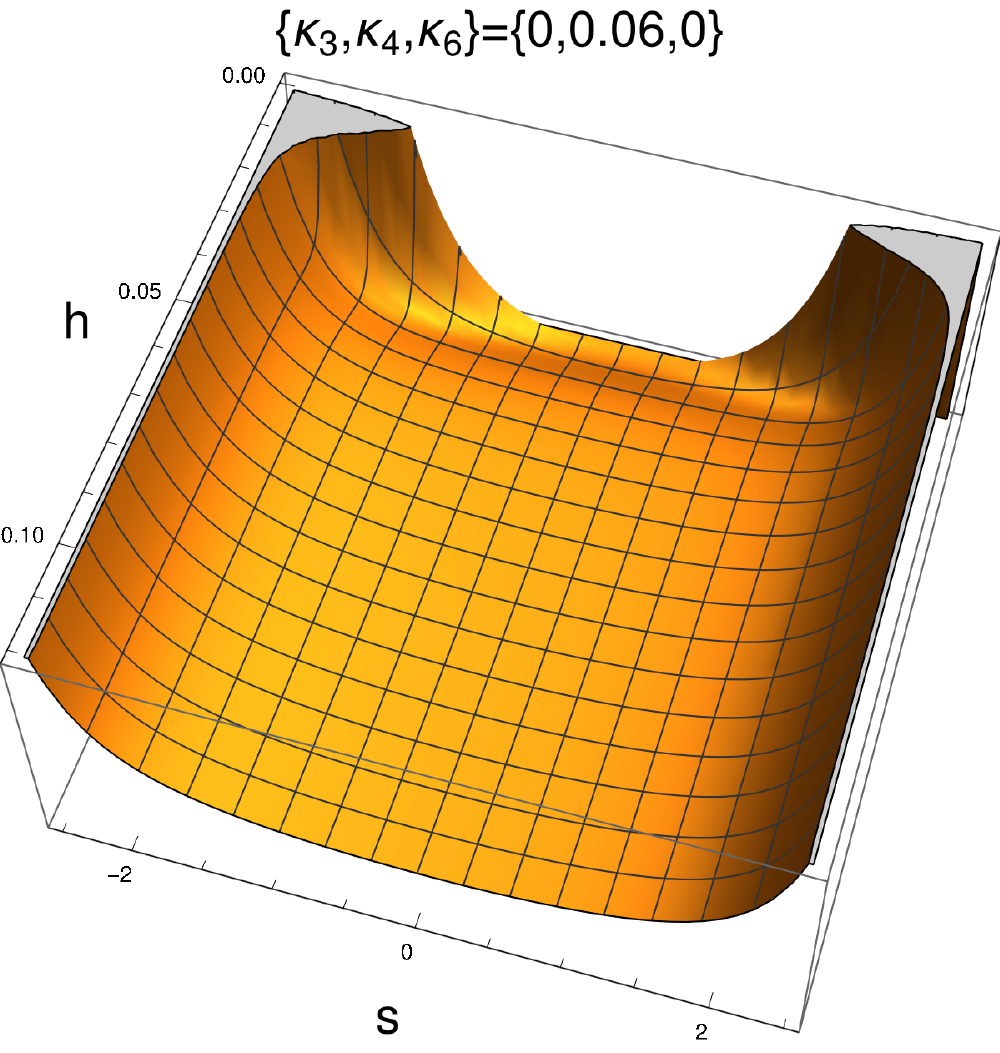}
\includegraphics[width=0.3\textwidth]{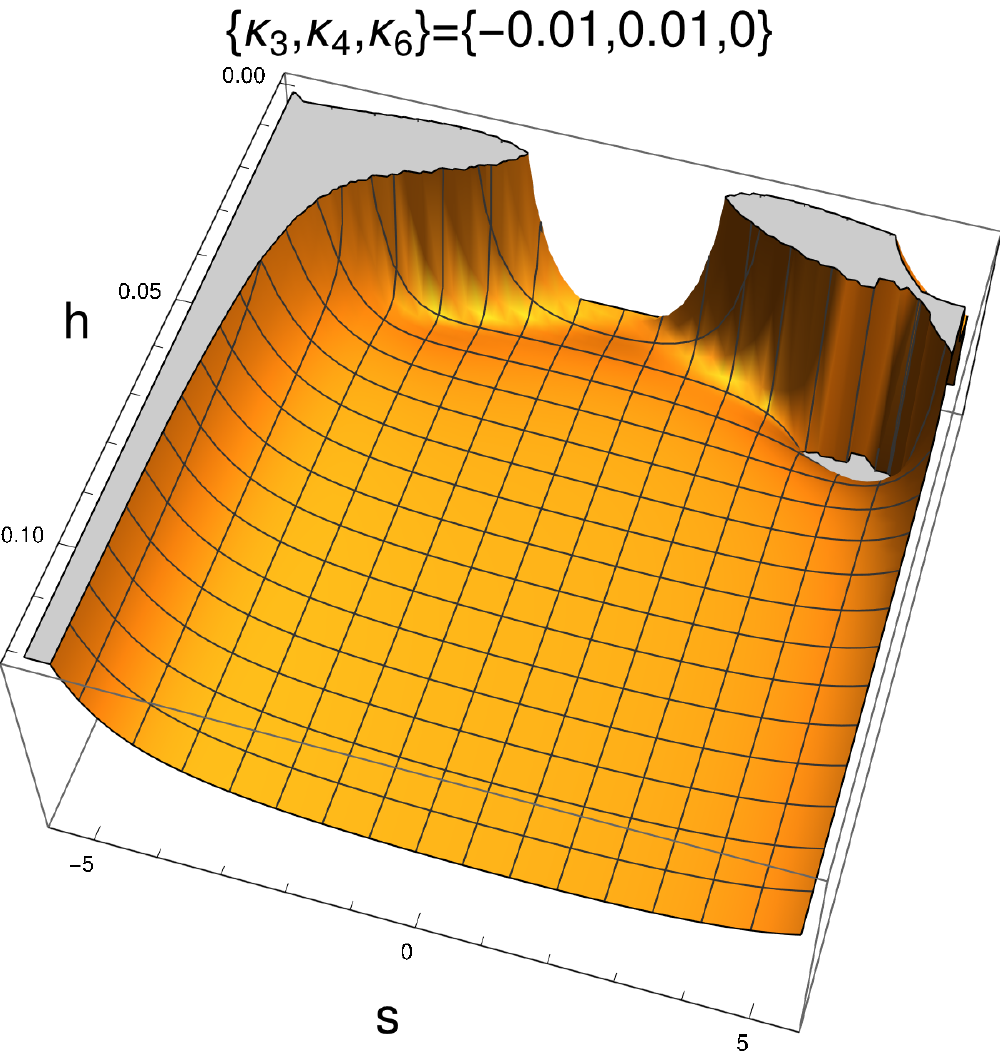} 
\caption{Typical shapes of the Einstein frame potential $V_{\rm E}$ for the cubic case $(\kappa_6=0)$. From left to right we have set $\{\kappa_3,\kappa_4\}=\{0,0.01\}$, $\{0,0.06\}$, and $\{-0.01,0.01\}$, with $\xi = 10^4$. For smaller values of $\kappa_4$ the potential develops two extra minima at $s\neq 0$. The $\kappa_3$ parameter introduces an asymmetry and tilts the scalar potential in the $s$-field direction.}
\label{fig:cubicpot}
\end{figure*}

\begin{figure*}[ht!]
\centering
\includegraphics[width=0.3\textwidth]{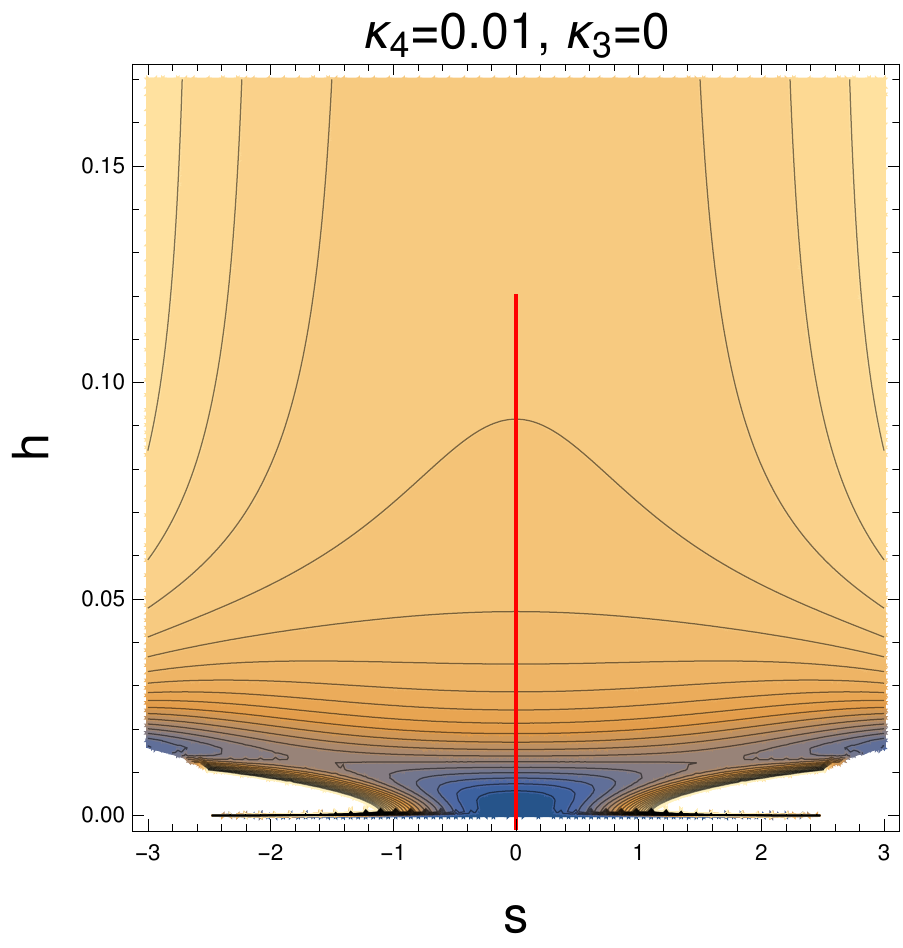} 
\includegraphics[width=0.3\textwidth]{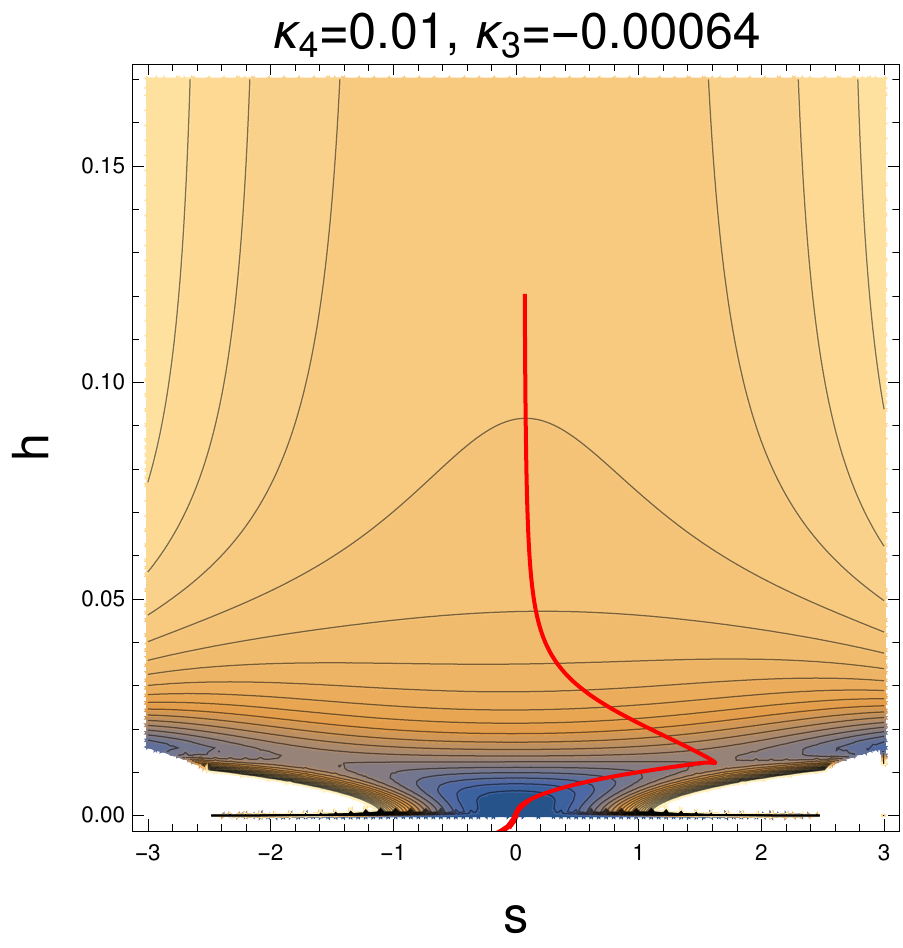}
\includegraphics[width=0.3\textwidth]{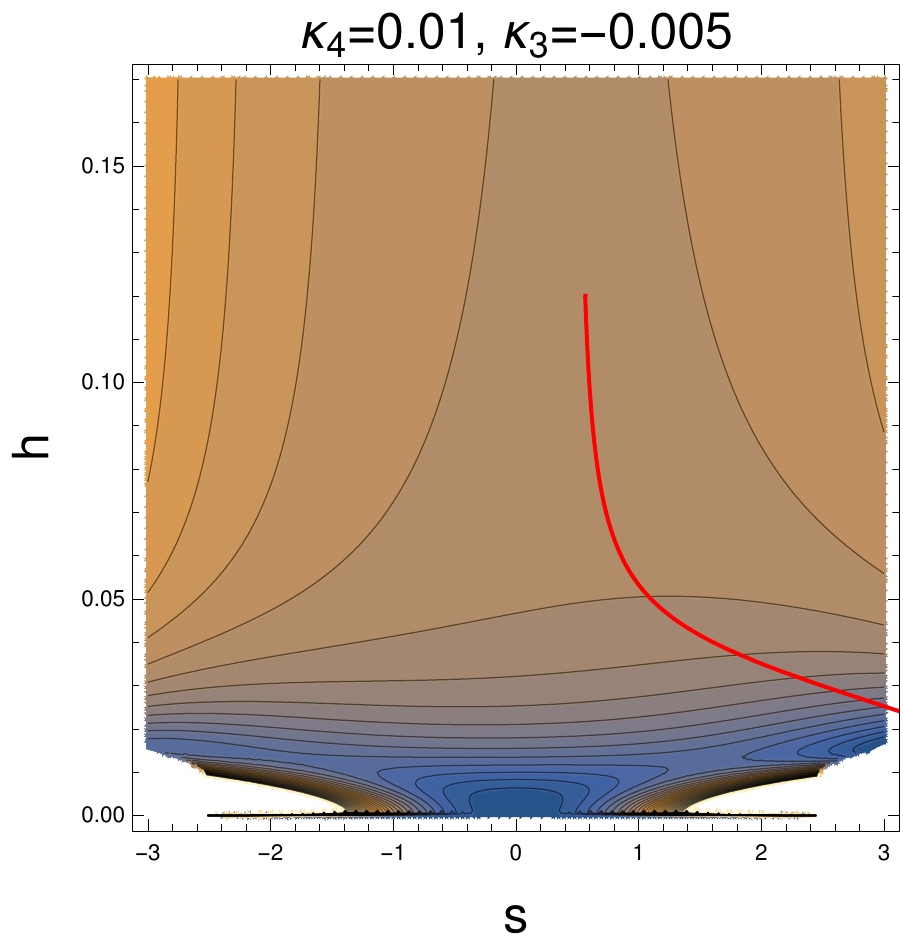} 
\caption{Examples of the inflaton trajectory for the cubic case $(\kappa_6=0)$. From left to right $\kappa_3 = 0$, $-0.00064$, and $-0.005$, with $\kappa_4 = 0.01$ and $\xi = 10^4$. The initial condition for the $h$ field is set to be 0.12. The initial $s$-field value is chosen to be at the potential minimum at $h=0.12$. The initial field velocities are fixed by using the slow roll background equations of motion. As $\kappa_3$ introduces a tilt in the scalar potential along the $s$-field direction, the inflaton trajectory deviates from a straight line. For a large negative $\kappa_3$ value, the trajectory falls into a false vacuum. With an appropriate value of $\kappa_3$, the inflaton trajectory may cross the saddle point between the false vacuum at $s\neq0$ and the true vacuum at $s=0$ as shown in the middle.}
\label{fig:cubictraj}
\end{figure*}

To ensure the stability of the scalar potential the quartic term needs to be positive, {\it i.e.}, $\kappa_4>0$.
Typical shapes of the Einstein frame potential \eqref{eqn:EFpot} are shown in Fig.~\ref{fig:cubicpot}.
One may see that, as the $\kappa_4$ parameter becomes smaller, two additional minima at $s\neq 0$ start to appear in the potential, as was pointed out in \cite{Ferrara:2010in}. Turning on $\kappa_3$ makes the scalar potential tilted to one side in the $s$-field direction, introducing an asymmetry.
Choosing an appropriate value for $\kappa_3$, we may adjust the inflaton trajectory in such a way that it passes the saddle point between the true vacuum at $s=0$ and the false vacuum\footnote{
Although somewhat misnamed, we refer to the minimum at $s=0$ as the `true vacuum' and the minimum at $s \neq 0$ as the `false vacuum,' regardless of the potential values at these points.
} at $s\neq 0$.

\begin{figure}[t!]
\centering
\includegraphics[scale=0.9]{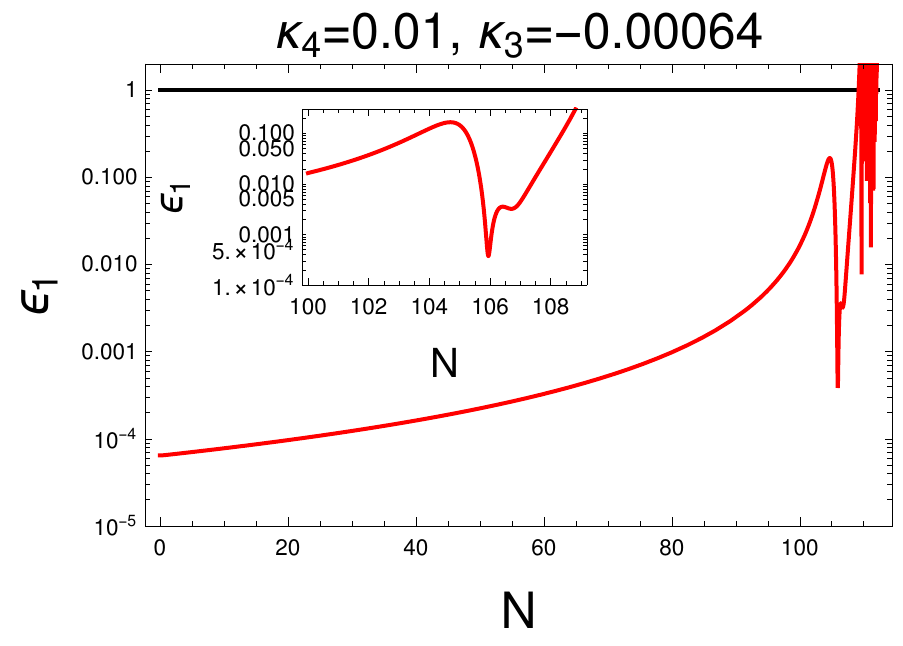}
\caption{First Hubble slow roll parameter for $\kappa_3=-0.00064$ and $\kappa_4=0.01$. As the inflaton trajectory crosses the saddle point $\epsilon_1$ gets suppressed, entering the ultra-slow roll regime.}
\label{fig:cubiceps1}
\end{figure}

We show three examples of the inflaton trajectory for different values of $\kappa_3$ in Fig.~\ref{fig:cubictraj}, fixing $\kappa_4 = 0.01$ and $\xi=10^4$.
When $\kappa_3=0$, we obtain a straight trajectory, while for a negatively large value of $\kappa_3$, the trajectory falls into a false vacuum at $s\neq0$\footnote{For positively large values of $\kappa_3$, the potential becomes tilted in the opposite direction, and thus the trajectory falls into a false vacuum at $s<0$. We do not consider this case.}. With an appropriate $\kappa_3$ value, the inflaton trajectory crosses the saddle point between the false vacuum and the true vacuum at $s=0$ as shown in the middle in Fig.~\ref{fig:cubictraj}.
When the trajectory crosses the saddle point, the first Hubble slow roll parameter $\epsilon_1 \equiv -\dot{H}/H^2$ gets suppressed as shown in Fig.~\ref{fig:cubiceps1}, indicating the ultra-slow roll regime. It hints at the possibility of curvature perturbation enhancement and thus PBH formations.

\begin{figure}[t!]
\centering
\includegraphics[scale=0.9]{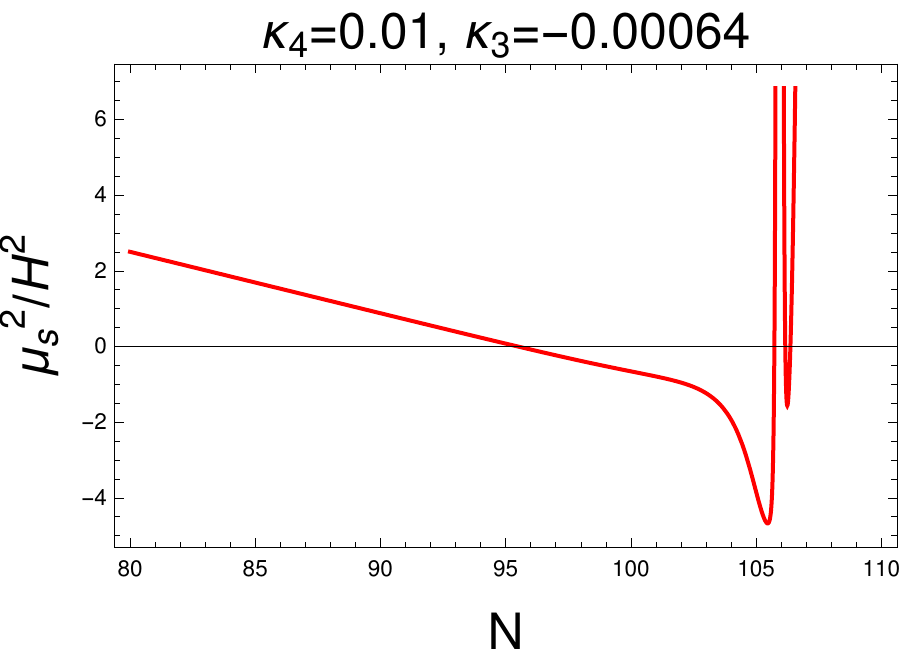}
\caption{Effective entropic mass squared for $\kappa_3=-0.00064$ and $\kappa_4=0.01$. As the inflaton trajectory crosses the saddle point the entropic mass squared becomes negative, {\it i.e.}, tachyonic.}
\label{fig:cubicmass}
\end{figure}

From the perturbation equation for the entropic perturbation \eqref{eqn:Qspert}, the effective entropic mass squared is defined as $\mu_s^2 \equiv \mathcal{M}_{ss} + 3\omega^2$. As the inflaton trajectory crosses the saddle point $\mu_s^2$ becomes negative as shown in Fig.~\ref{fig:cubicmass}. In other words the entropic perturbation briefly enters the tachyonic regime, sourcing the growth of the perturbation. The growth of the entropic perturbation in turn may source the curvature perturbation. This is a unique feature of multifield inflation.

\begin{figure*}[ht!]
\centering
\includegraphics[width=0.3\textwidth]{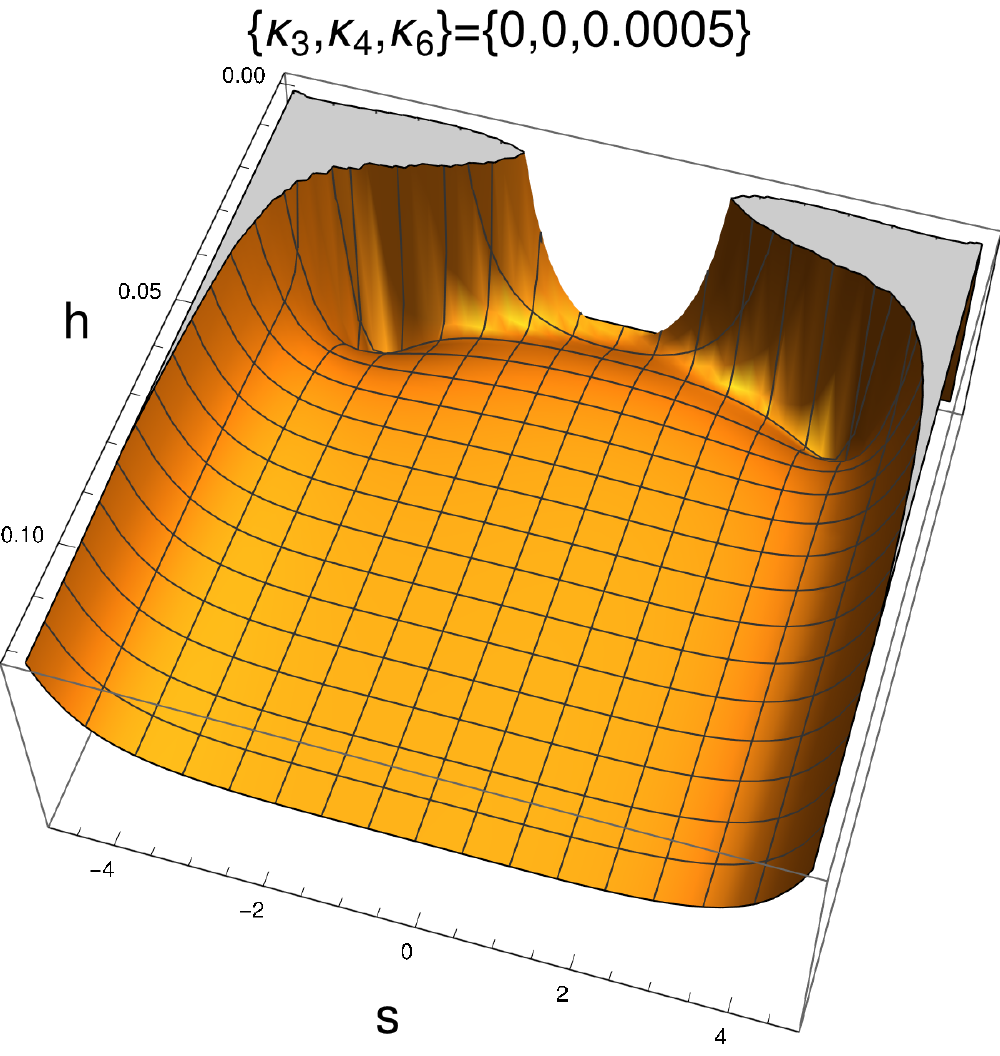} 
\includegraphics[width=0.3\textwidth]{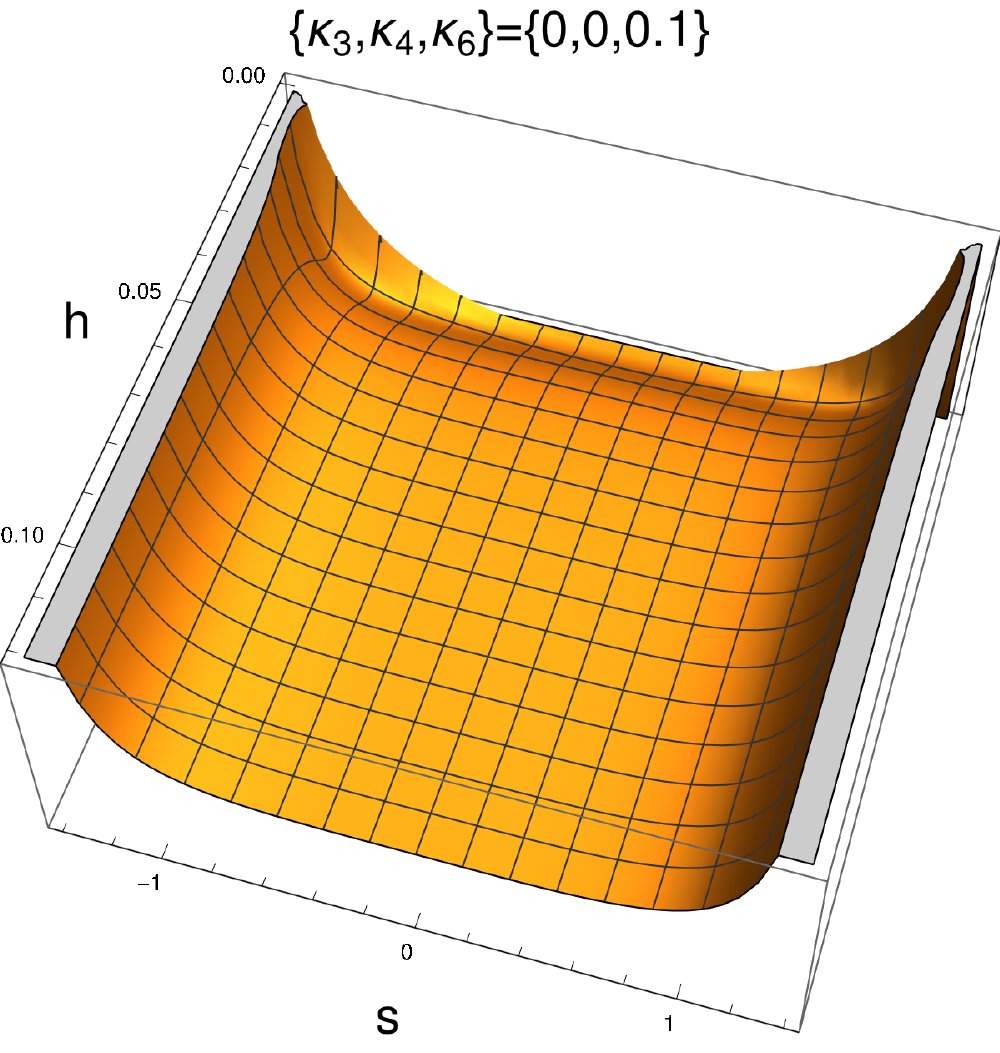}
\includegraphics[width=0.3\textwidth]{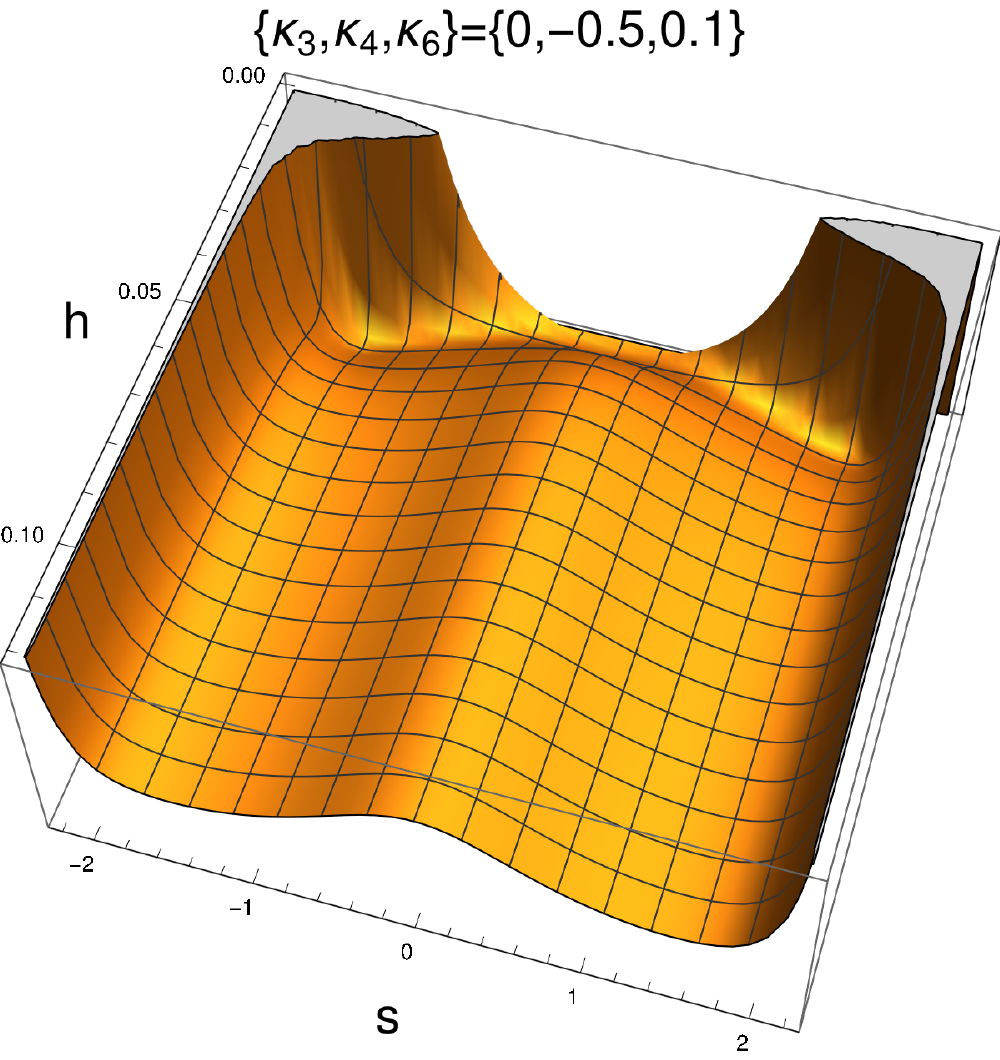} 
\caption{Typical shapes of the Einstein frame potential $V_{\rm E}$ for the sextic case ($\kappa_3=0$). From left to right we have set $\{\kappa_4,\kappa_6\}=\{0,0.0005\}$, $\{0,0.1\}$, and $\{-0.5,0.1\}$, with $\xi=10^4$.}
\label{fig:sexticpot}
\end{figure*}

\begin{figure*}[ht!]
\centering
\includegraphics[width=0.23\textwidth]{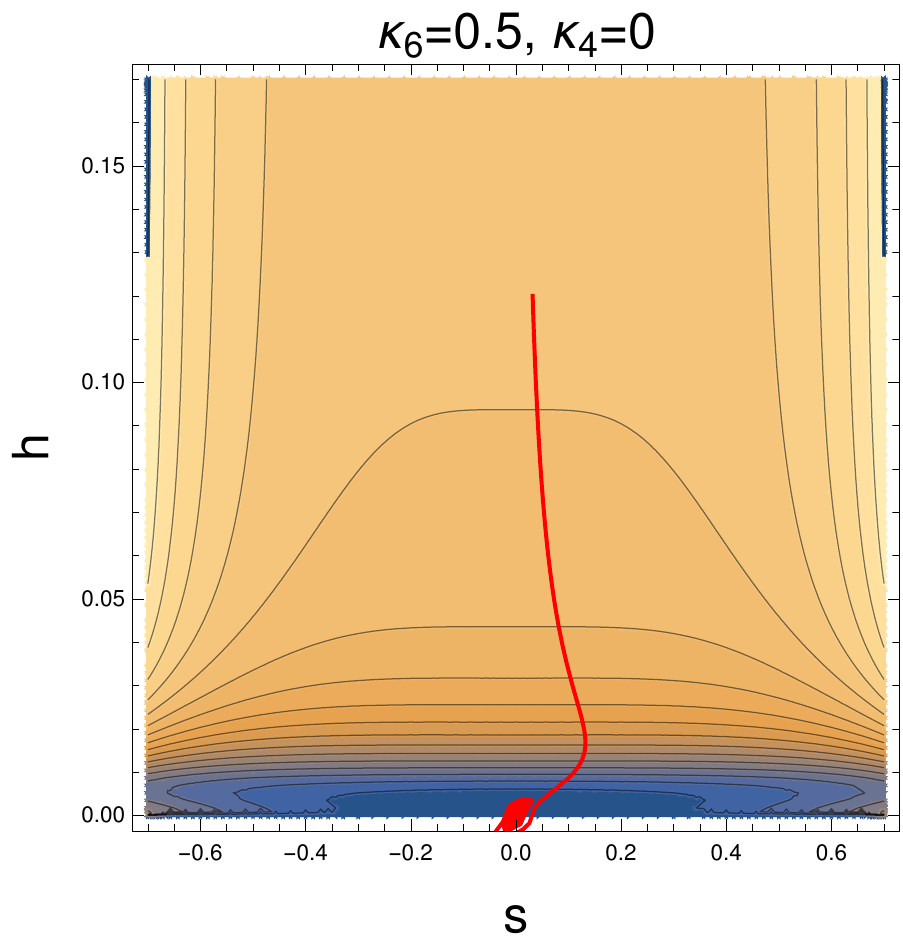} 
\includegraphics[width=0.23\textwidth]{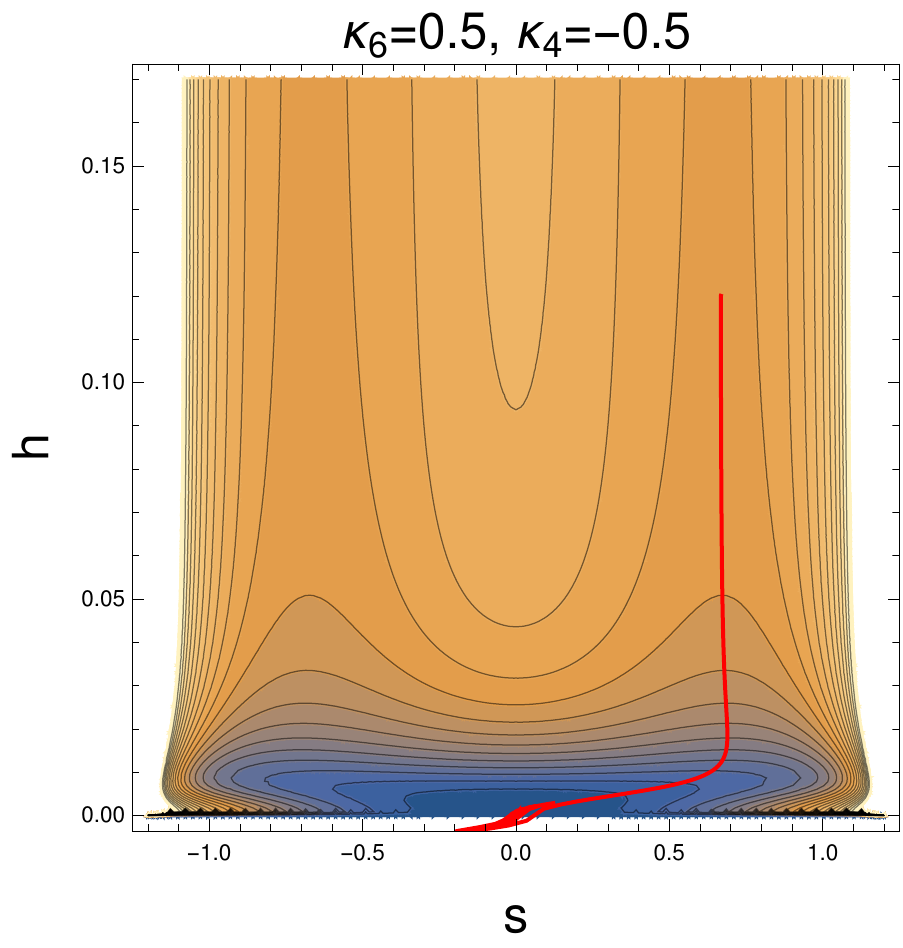}
\includegraphics[width=0.23\textwidth]{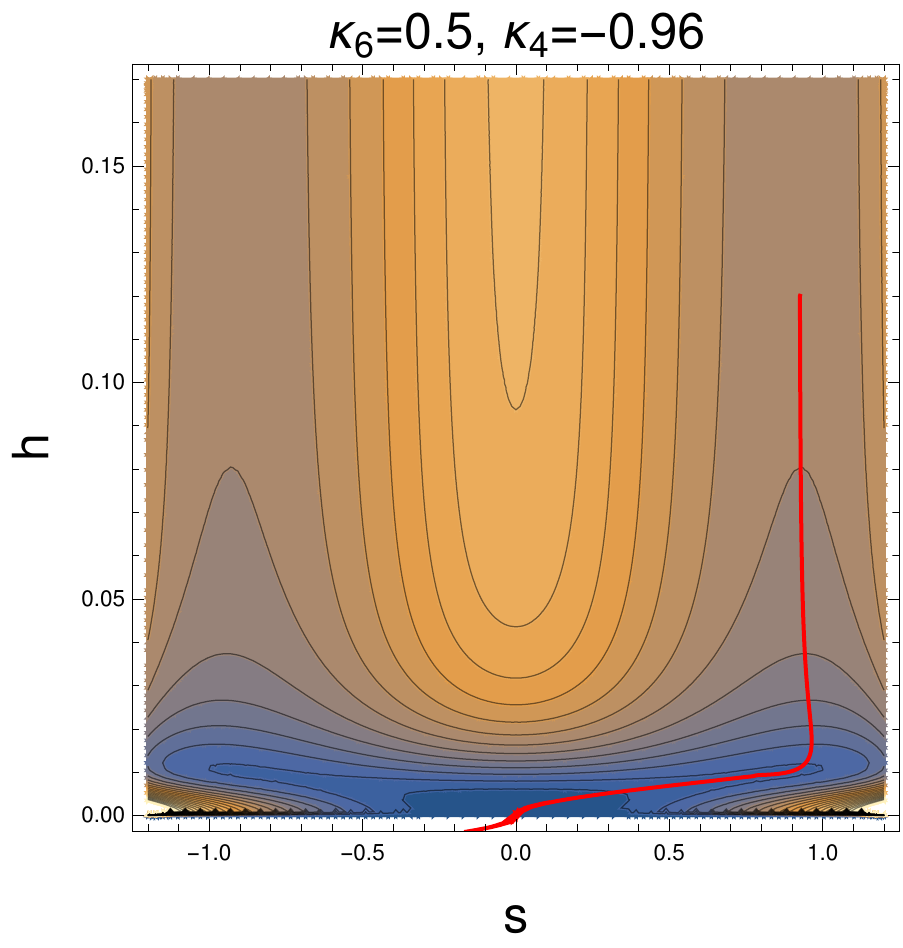}
\includegraphics[width=0.23\textwidth]{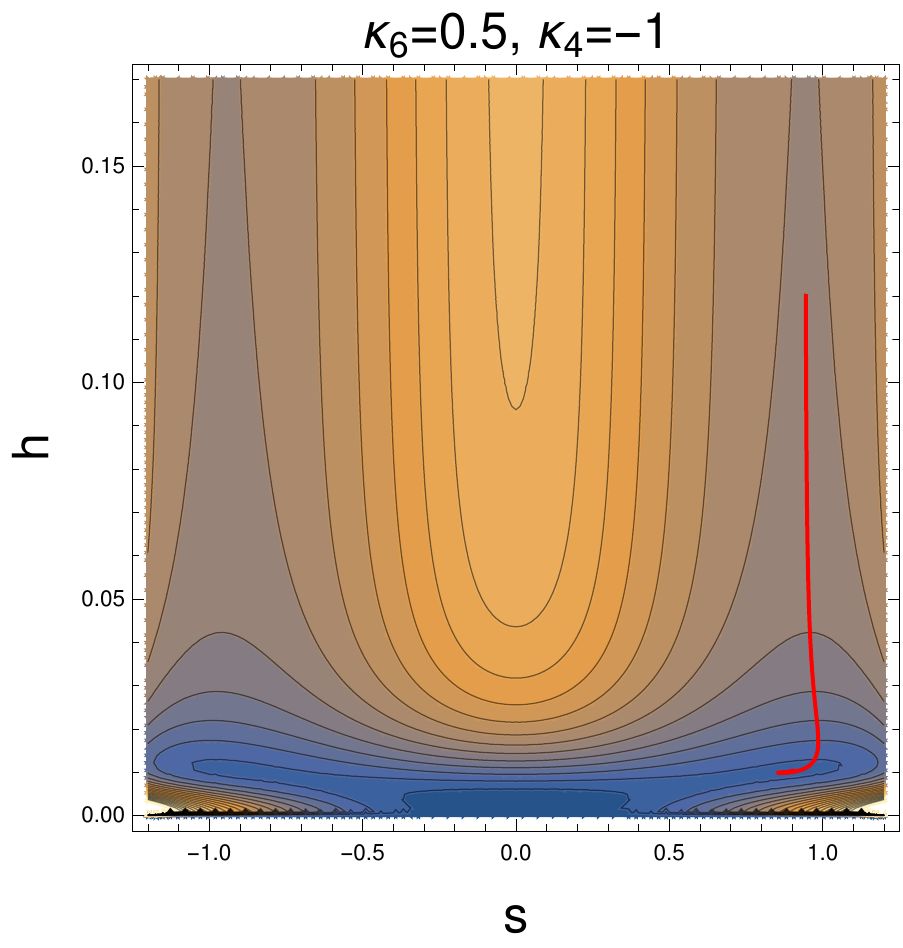} 
\caption{Examples of the inflaton trajectory for the sextic case $(\kappa_3=0)$. From left to right $\kappa_4 = 0$, $-0.5$, $-0.96$, and $-1$, with $\kappa_6 = 0.5$ and $\xi = 10^4$. The initial condition for the $h$ field is set to be 0.12. The initial $s$-field value is chosen to be at the potential minimum at $h=0.12$. The initial field velocities are fixed by using the slow roll background equations of motion. Note that, unlike the cubic case, $s=0$ is no longer a stable position at $h_{\rm ini}$. As $\kappa_4$ negatively increases, the false vacuum at $s\neq0$ becomes deeper than the true vacuum at $s=0$; for example, we see that the inflaton trajectory gets trapped at the false vacuum when $\kappa_4=-1$. With an appropriate $\kappa_4$ value, one may balance between the false vacuum and the true vacuum, generating an ultra-slow roll regime.}
\label{fig:sextictraj}
\end{figure*}

\begin{figure}[t!]
\centering
\includegraphics[scale=0.9]{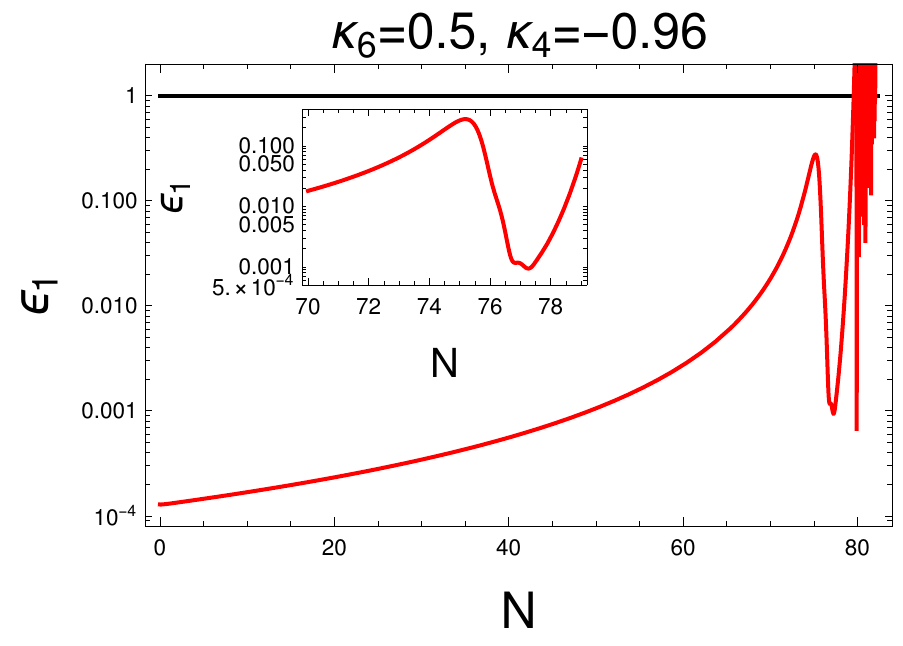}
\caption{First Hubble slow roll parameter for $\kappa_6=0.5$ and $\kappa_4=-0.96$. The suppression of $\epsilon_1$ indicates the curvature power spectrum enhancement and thus PBH formations.}
\label{fig:sexticeps1}
\end{figure}

\subsection{Sextic case}
In the sextic case, $\kappa_6>0$ is required for the stability of the scalar potential.
Figure~\ref{fig:sexticpot} shows typical Einstein frame potential shapes. Note that, unlike the cubic case, $s=0$ is not a stable point at $h=h_{\rm ini}$. Similar to the cubic case, when $\kappa_4=0$, smaller values of $\kappa_6$ introduce false vacua at $s\neq0$. The role of the $\kappa_4$ parameter is two-fold. One is to shift the potential minimum in the large-$h$ limit towards a larger $s$-field value. The other is to create or deepen false vacua at $s\neq0$.

We present four samples of the inflaton trajectory for different values of $\kappa_4$ in Fig.~\ref{fig:sextictraj} with $\kappa_6=0.5$ and $\xi=10^4$. One can clearly see the effect of $\kappa_4$; once $\kappa_4$ takes a negatively large value, the inflaton trajectory gets trapped in the false vacuum. Adjusting the value of $\kappa_4$, we may balance the heights of the false vacuum and the true vacuum. In this case the inflaton undergoes an ultra-slow roll regime. This case demonstrated in Fig.~\ref{fig:sexticeps1} which shows the first Hubble slow roll parameter $\epsilon_1$. Similar to the cubic case we observe the suppression of $\epsilon_1$ near the ultra-slow roll regime which may lead to the enhancement of the curvature perturbation. Moreover the tachyonic behavior is also seen in the sextic case although the degree is less significant compared to the cubic case. Thus the sextic case as well has the possibility of PBH formations.

\begin{figure*}[ht!]
\centering
\includegraphics[width=0.45\textwidth]{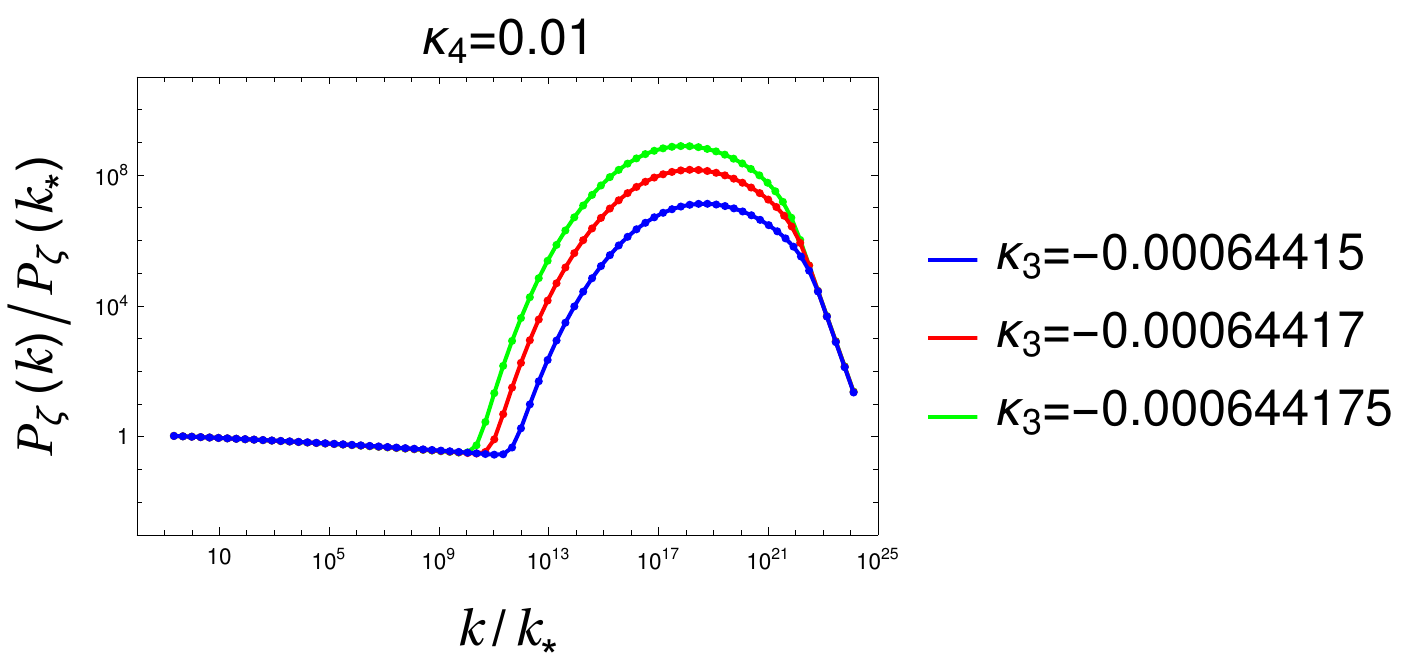} 
\includegraphics[width=0.45\textwidth]{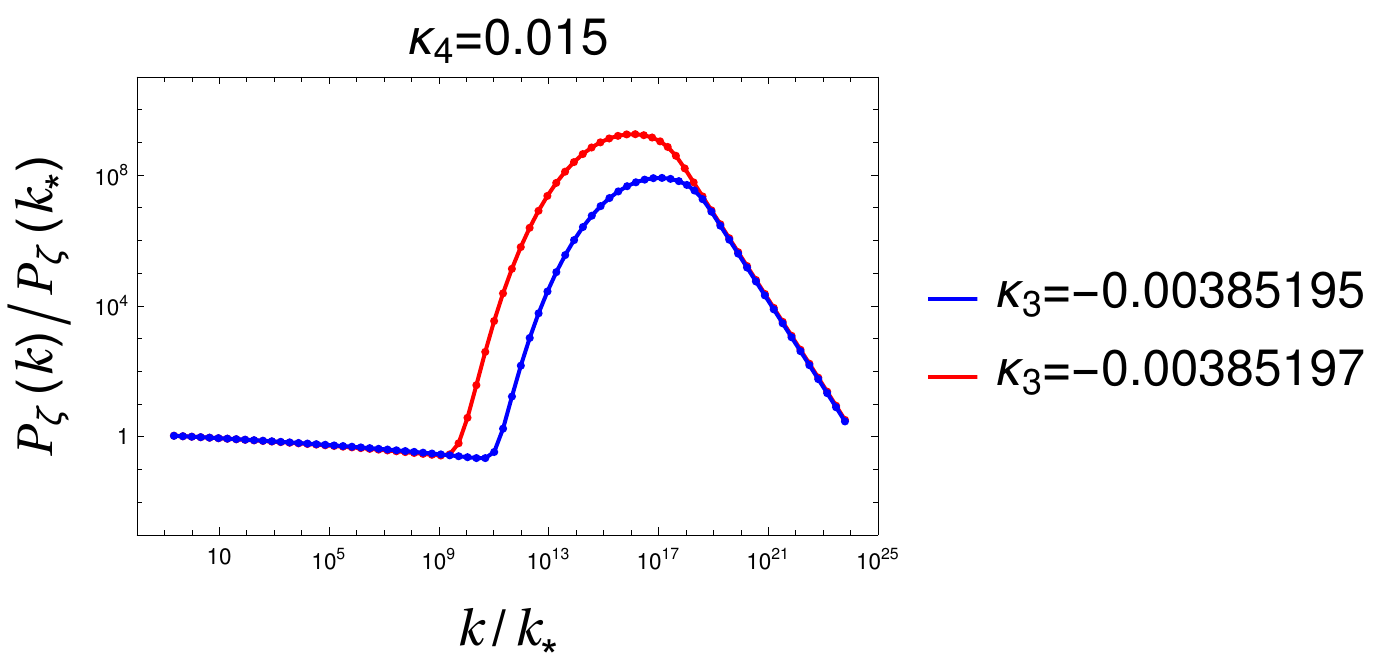} \\
\includegraphics[width=0.45\textwidth]{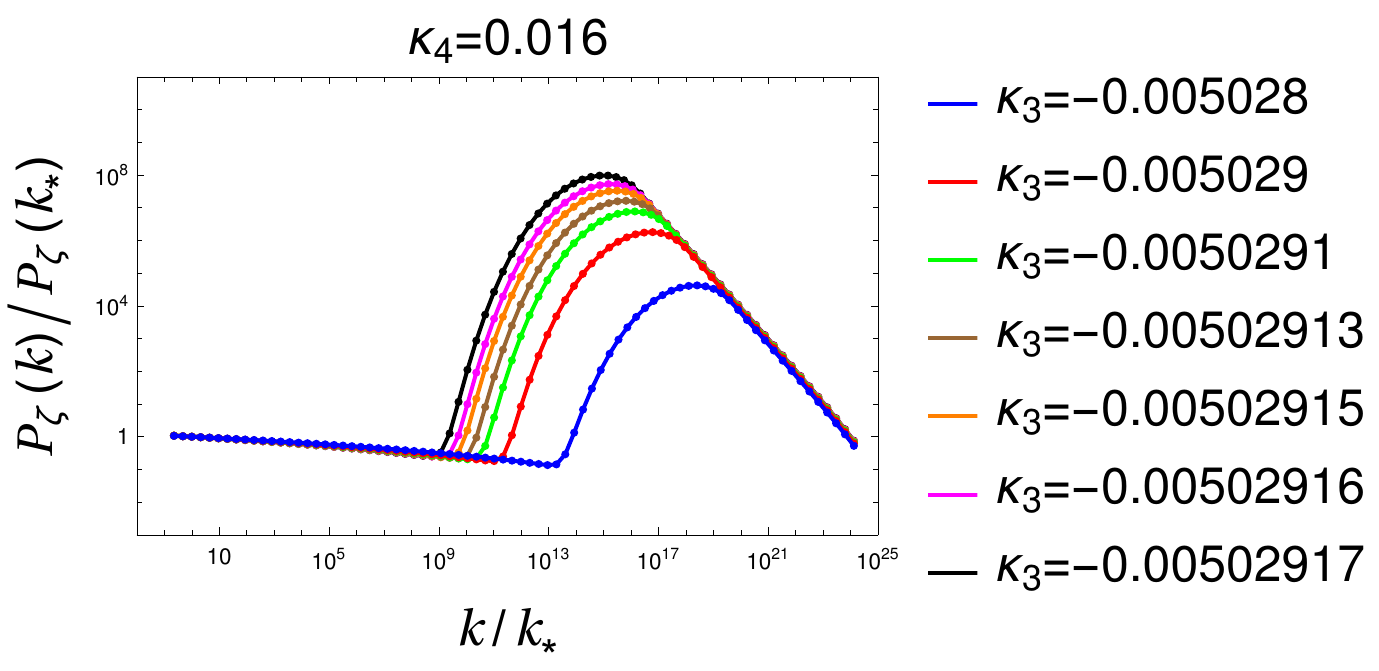} 
\includegraphics[width=0.45\textwidth]{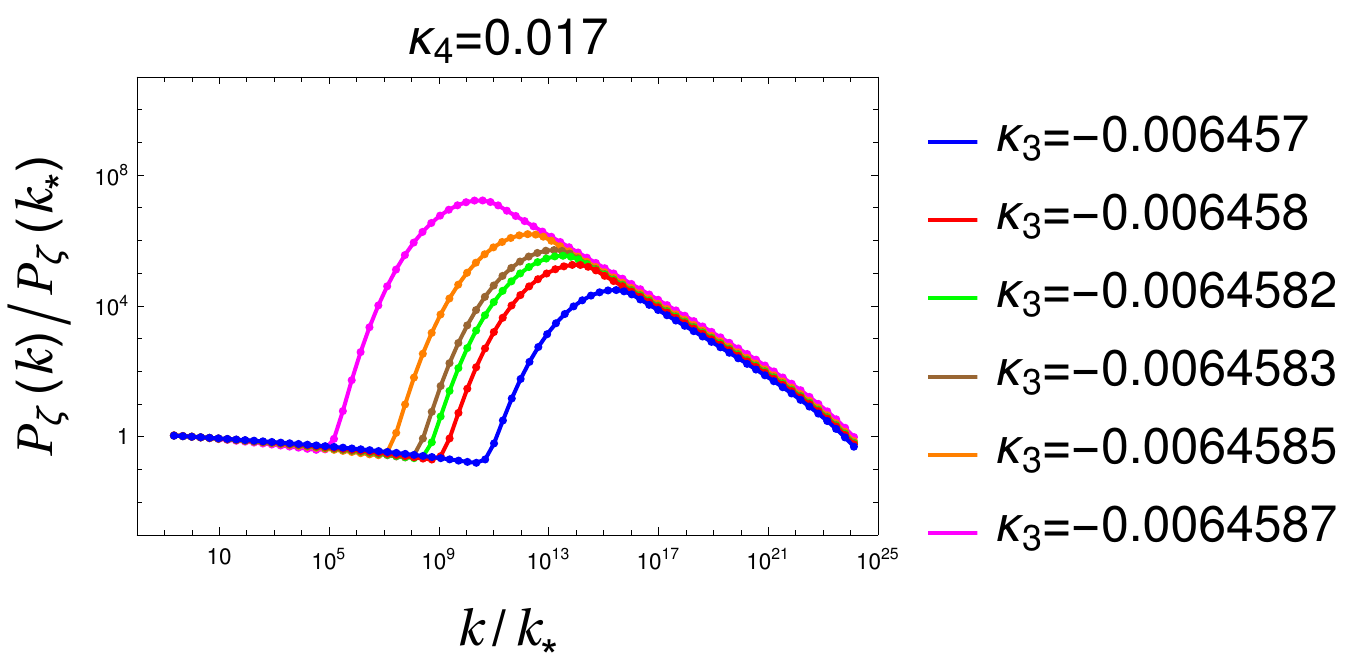} 
\caption{Curvature power spectrum for the cubic case. For a fixed value of $\kappa_4$, the peak position and the magnitude of the enhancement are controlled by the $\kappa_3$ parameter. For a fixed value of the curvature power spectrum, the peak position is controlled by the $\kappa_4$ parameter. As $\kappa_4$ increases, the peak position moves towards a lower wavenumber.}
\label{fig:cubicCPS}
\end{figure*}

\section{Curvature Power Spectrum}
\label{sec:curvPS}
The suppression of the slow roll parameter $\epsilon_1$ suggests the possibility of curvature perturbation enhancement. In this section we demonstrate that this is indeed the case. To compute the curvature power spectrum $\mathcal{P}_\zeta(k)$ we adopt the transport method\footnote{
Packages implementing the transport method are publicly available, including {\tt mTransport} \cite{Dias:2015rca}, {\tt CppTransport} \cite{Dias:2016rjq,Seery:2016lko}, {\tt PyTransport} \cite{Dias:2016rjq,Mulryne:2016mzv,Ronayne:2017qzn}, and {\tt Inflation.jl} \cite{Rosati:2020ijl}.
} \cite{Mulryne:2009kh,Mulryne:2010rp,Dias:2011xy,Anderson:2012em,Elliston:2012ab,Mulryne:2013uka,Dias:2014msa,Dias:2015rca,Dias:2016rjq}, which, with the assumption of instantaneous reheating and small isocurvature mode, evolves the system from deep inside the horizon\footnote{
We have chosen 8 $e$-foldings before the horizon exit and checked convergence for larger values.
} until the end of inflation specified by the condition $\epsilon_1 = 1$.

In Fig.~\ref{fig:cubicCPS} the curvature power spectrum, normalized by its value at the pivot scale, is shown as a function of the wavenumber for different values of $\kappa_3$ and $\kappa_4$ for the cubic case. For all the cases we have set $\xi=10^4$. We can see that, for a fixed value of $\kappa_4$, the $\kappa_3$ parameter controls the magnitude as well as the peak position of the enhancement. For a fixed value of the curvature power spectrum, say $\mathcal{P}_\zeta(k)/\mathcal{P}_\zeta(k_*) = 10^6$, the $\kappa_4$ parameter controls the position of the peak; the peak position moves towards a smaller wavenumber as $\kappa_4$ increases.

\begin{figure*}[ht!]
\centering
\includegraphics[width=0.45\textwidth]{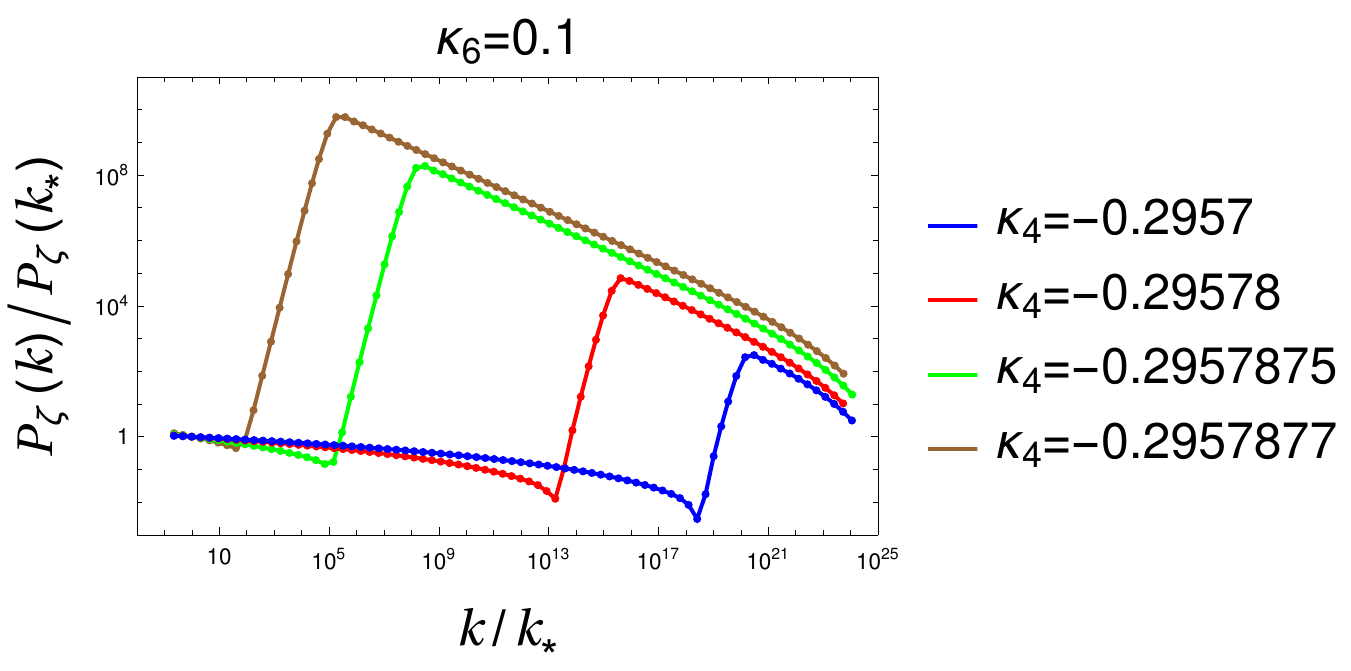} 
\includegraphics[width=0.45\textwidth]{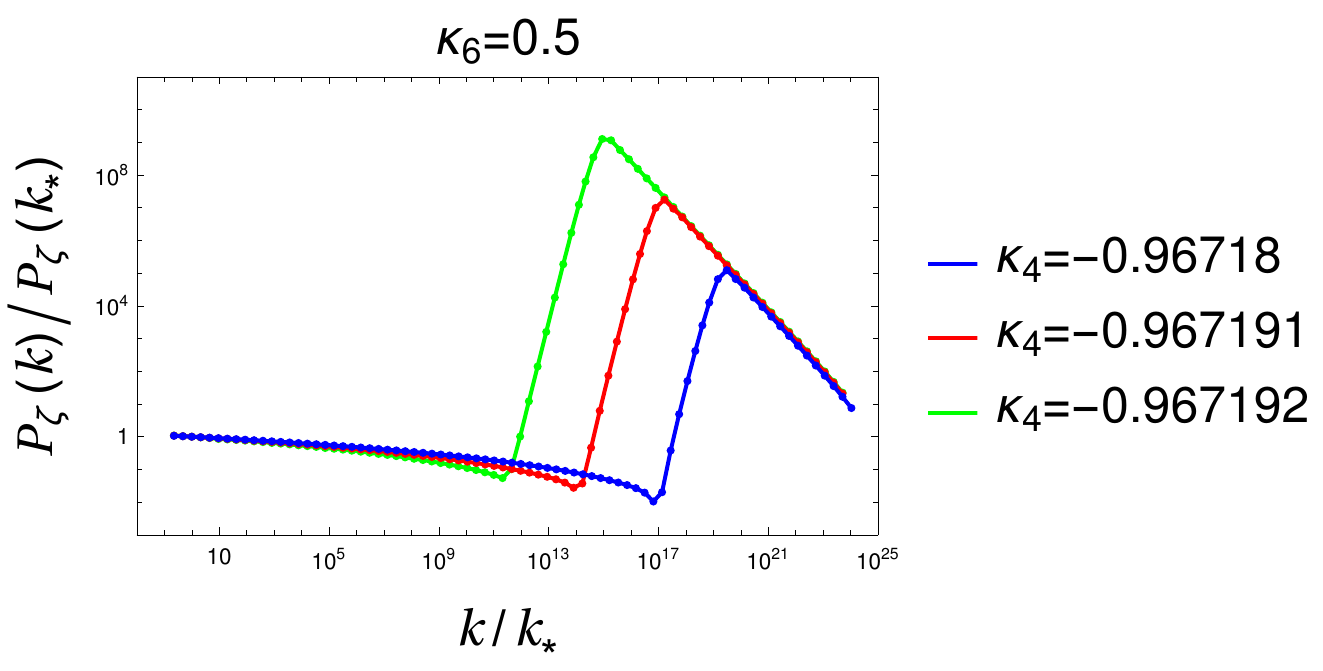} \\
\includegraphics[width=0.45\textwidth]{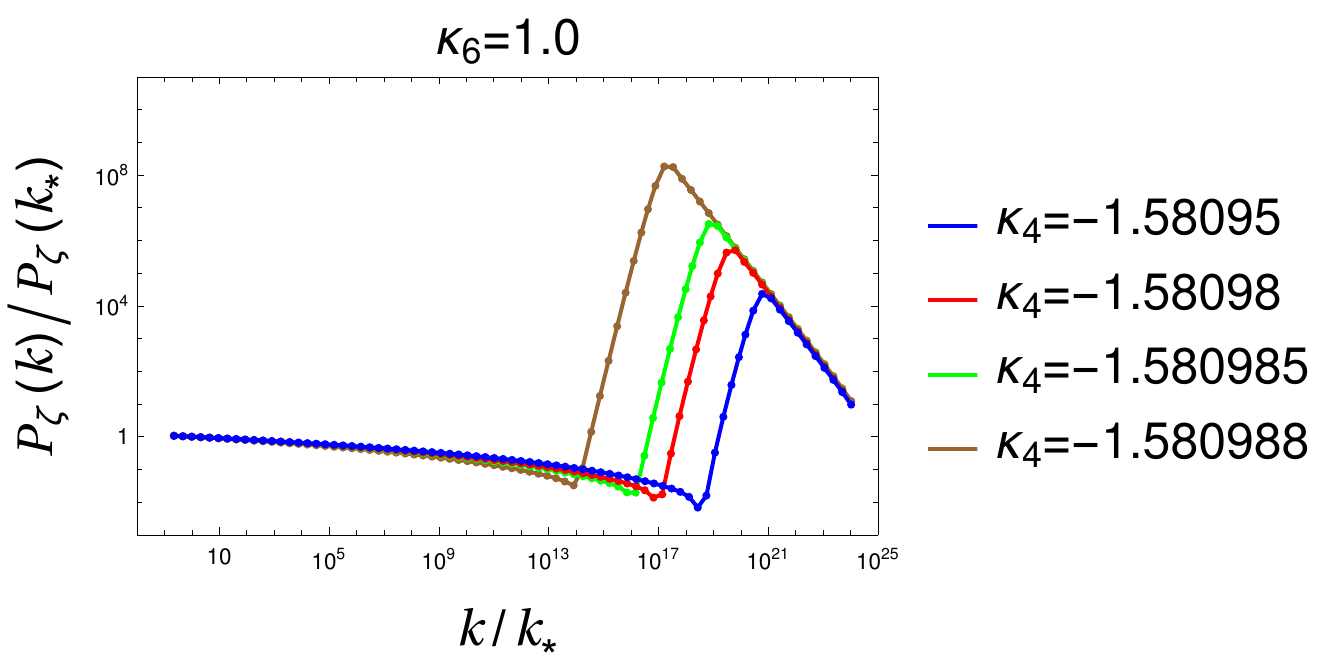} 
\includegraphics[width=0.45\textwidth]{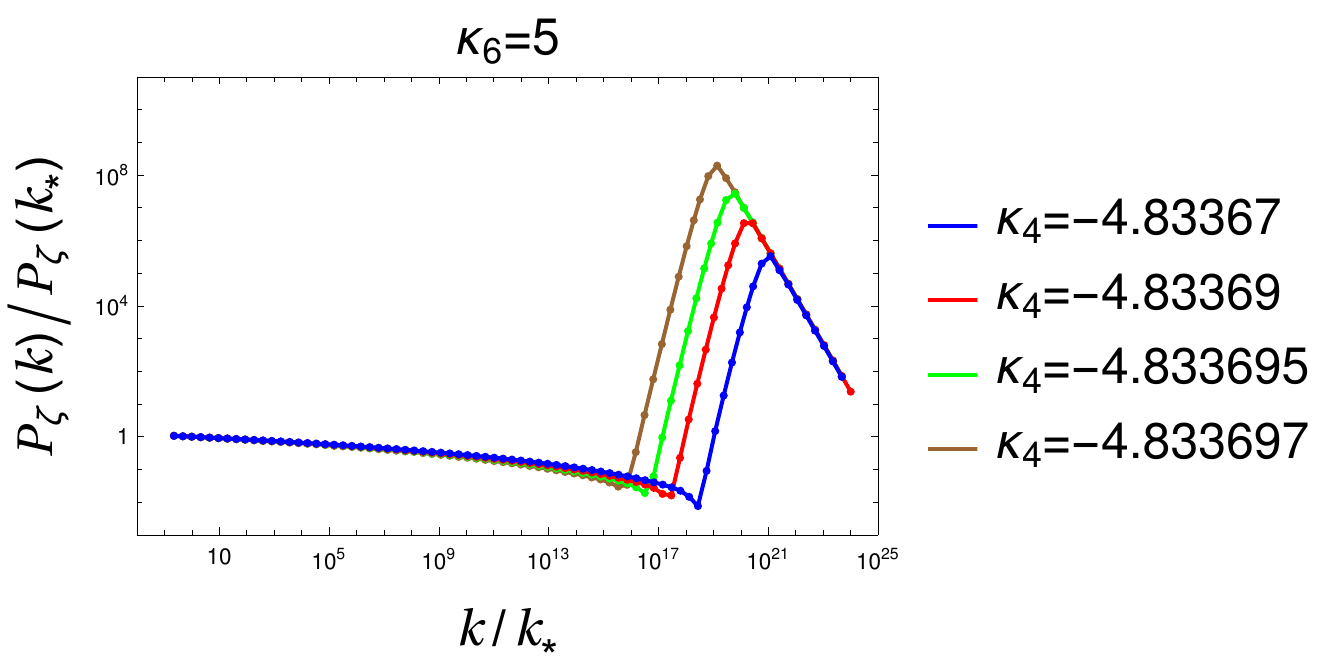} 
\caption{Curvature power spectrum for the sextic case. For a fixed value of $\kappa_6$, the peak position and the magnitude of the enhancement are controlled by the $\kappa_4$ parameter. For a fixed value of the curvature power spectrum, the peak position is controlled by the $\kappa_6$ parameter. As $\kappa_6$ increases, the peak position moves towards a higher wavenumber.}
\label{fig:sexticCPS}
\end{figure*}

Figure~\ref{fig:sexticCPS} shows the curvature power spectrum, normalized by its value at the pivot scale as a function of the wavenumber for different values of $\kappa_4$ and $\kappa_6$ for the sextic case. We have again set $\xi=10^4$. Similar to the cubic case, the peak position and the magnitude of the enhancement are controlled by the $\kappa_4$ parameter for a fixed value of $\kappa_6$. On the other words, the $\kappa_6$ parameter is the key controller of the peak position for a fixed value of the curvature power spectrum; in this case, unlike the cubic case, the peak position moves towards a larger wavenumber as $\kappa_6$ increases.

\section{Prediction of the model and future detectability}
\label{sec:PBHsGWs}
As a consequence of the curvature perturbation enhancement, PBHs may form through the gravitational collapse at the horizon reentry and constitute a large part of the today's dark matter relic density.
Another consequence is the generation of scalar-sourced stochastic gravitational waves at the nonlinear order.
In this section we discuss the prediction of the model on the formation of PBHs and the generation of the gravitational waves.
Aligning with the assumption of instantaneous reheating, we consider that the PBH formation and the generation of the gravitational waves happen during the radiation-dominated era
\footnote{
The effects of non-instantaneous reheating, which are model-dependent, can be absorbed by a slight shift in the number of $e$-folds and do not alter the conclusion of our paper; see also footnote 3.
Note further that the relative scale between the CMB and the mode corresponding to PBHs and gravitation waves are insensitive to the reheating process.
}.

\subsection{Primordial Black Holes}
The produced PBHs have the mass of
\begin{align}
M = \gamma M_{H,0}\Omega_{{\rm rad}, 0}^{1/2}\left(
\frac{g_{*,0}}{g_{*,{\rm f}}}
\right)^{1/6}\left(
\frac{k_0}{k_{\rm f}}
\right)^2\,,
\end{align}
at the formation time, where $M_{H,0}=4\pi/H$ is the horizon mass, $\Omega_{\rm rad}$ is the radiation energy density parameter, $g_*$ is the effective degrees of freedom, the subscript ${\rm f}$ ($0$) denotes the formation time (today), and the factor $\gamma$ describes the fraction of the horizon mass that turns into the PBHs. Following simple analytical estimations \cite{Carr:1975qj} we use $\gamma = 0.2$.

The energy density of the PBHs today can be obtained by redshifting that at the formation time, namely $\rho_{{\rm PBH}, 0} = \rho_{{\rm PBH}, {\rm f}}(a_{\rm f}/a_0)^3 \approx \gamma \beta \rho_{{\rm rad},{\rm f}}(a_{\rm f}/a_0)^3$, since the PBHs behave as matter. Here $\beta$ denotes the probability of the density fluctuation $\delta>\delta_c$, which is given by
\begin{align}
\beta = \int_{\delta_c} d\delta \,
\frac{1}{\sqrt{2\pi\sigma^2}}\exp\left(
-\frac{\delta^2}{2\sigma^2}
\right)\,,
\end{align}
where $\delta_c = 1/3$ \cite{Carr:1975qj} is a threshold value and $\sigma^2$ is the variance \cite{Josan:2009qn,Young:2014ana}
\begin{align}
\sigma^2 = \frac{16}{81}
\int_0^\infty 
\frac{dq}{q}\,\left(
\frac{q}{k}
\right)^4
W^2\left(
\frac{q}{k}
\right)
\mathcal{P}_\zeta (q)\,.
\end{align}
Note that we have assumed a Gaussian density fluctuation, neglecting possible effects of non-Gaussianities \cite{Bullock:1996at,Ivanov:1997ia,Lyth:2012yp,Byrnes:2012yx,Bugaev:2013vba,Young:2013oia,Nakama:2016gzw,Garcia-Bellido:2017aan,Franciolini:2018vbk,Cai:2018dig,Atal:2018neu,Unal:2018yaa,Passaglia:2018ixg,Atal:2019cdz,Panagopoulos:2019ail,Yoo:2019pma,Kehagias:2019eil,Ezquiaga:2019ftu,Yuan:2020iwf,Ragavendra:2020sop,Adshead:2021hnm,Atal:2021jyo}, and we take the Gaussian function for the window function $W(q/k)=\exp(-(q/k)^2/2)$.

The total abundance is $\Omega_{{\rm PBH},{\rm tot}} = \int d\ln M \Omega_{\rm PBH}$, with $\Omega_{\rm PBH}$ expressed in terms of $f_{\rm PBH}$ which is given by
\begin{align}
f_{\rm PBH} &\equiv 
\frac{\Omega_{{\rm PBH},0}}{\Omega_{{\rm CDM},0}}
\nonumber\\
&\approx 
\left(
\frac{\beta}{3.27\times 10^{-8}}
\right)\left(
\frac{\gamma}{0.2}
\right)^\frac{3}{2}\left(
\frac{106.75}{g_{*,{\rm f}}}
\right)^\frac{1}{4}
\nonumber\\
&\times\left(
\frac{0.12}{\Omega_{{\rm CDM},0}h^2}
\right)\left(
\frac{M}{M_\odot}
\right)^{-\frac{1}{2}}\,,
\end{align}
where $\Omega_{{\rm CDM},0}$ is the today's density parameter of the cold dark matter, $h$ is the rescaled Hubble rate today, and $M_\odot$ is the solar mass.

\begin{figure*}[ht!]
\centering
\includegraphics[width=0.495\textwidth]{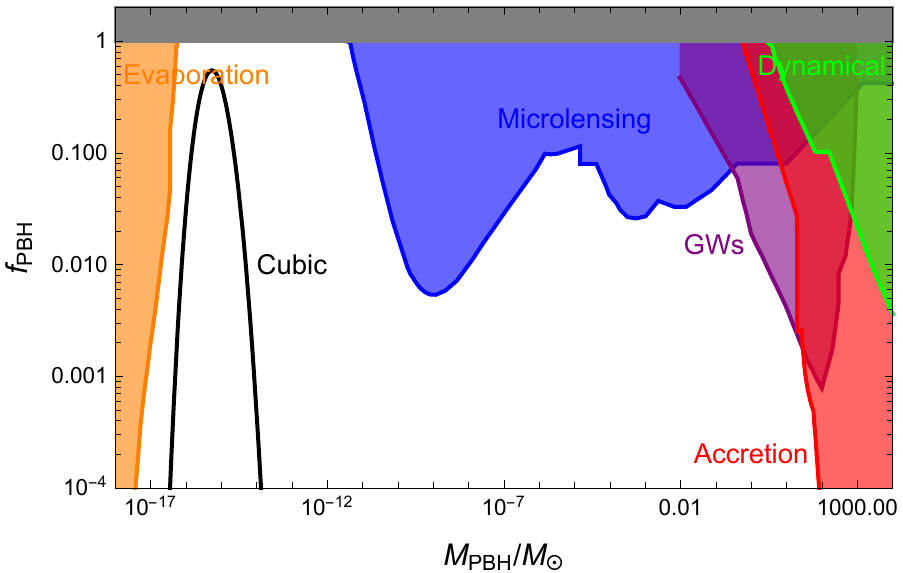}
\includegraphics[width=0.495\textwidth]{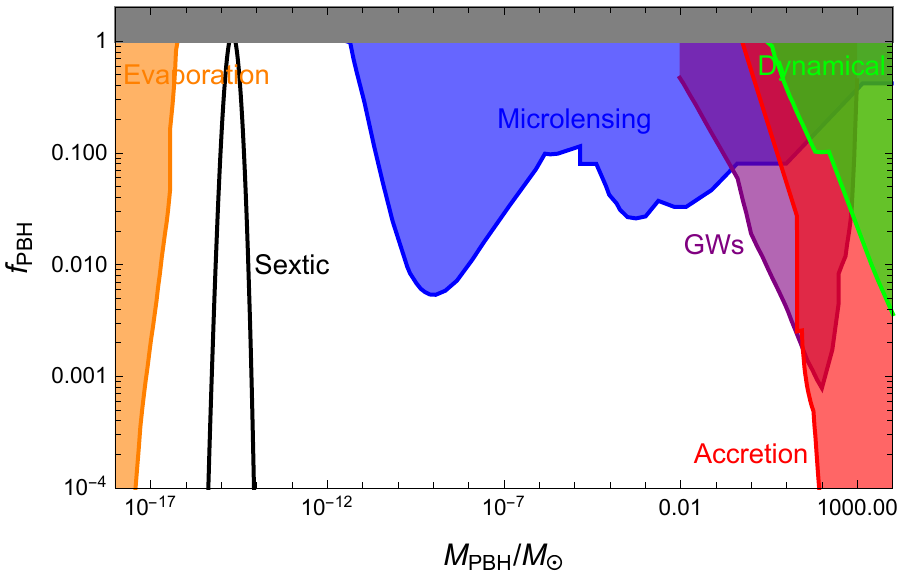}
\caption{The spectrum of the produced PBHs.
The left panel shows the case for the cubic model, where $\kappa_4 = 0.0163$ is chosen, and $\kappa_3$ is tuned to match $f_{\rm PBH}^{\rm tot} \approx 1$.
The right panel is the sextic model, where $\kappa_6 = 0.2$ is chosen, and $\kappa_4$ is tuned to match $f_{\rm PBH}^{\rm tot} \approx 1$.
We consulted \cite{Green:2020jor,Kavanagh:2019aaa} for the constraints data. See also Refs.~\cite{Laha:2019ssq,Laha:2020ivk,Saha:2021pqf} for additional constraints such as those coming from gamma-ray observations and the global 21cm signal, and Ref.~\cite{Ray:2021mxu} for forecasted constraints from future gamma-ray telescopes.}
\label{fig:pbh}
\end{figure*}

In Fig.~\ref{fig:pbh} we present two benchmark points for the PBH spectrum for the cubic and sextic cases. For the cubic case $\kappa_4 = 0.0163$ is chosen and $\kappa_3$ is tuned to match $f_{\rm PBH}^{\rm tot}$ to be unity. For the sextic model we chose $\kappa_6 = 0.2$, and $\kappa_4$ is tuned so that $f_{\rm PBH}^{\rm tot}$ becomes unity. Thus both the cubic and sextic models may produce enough PBHs to constitute all of the current dark matter abundance.

\begin{figure*}[ht]
\centering
\includegraphics[scale=0.85]{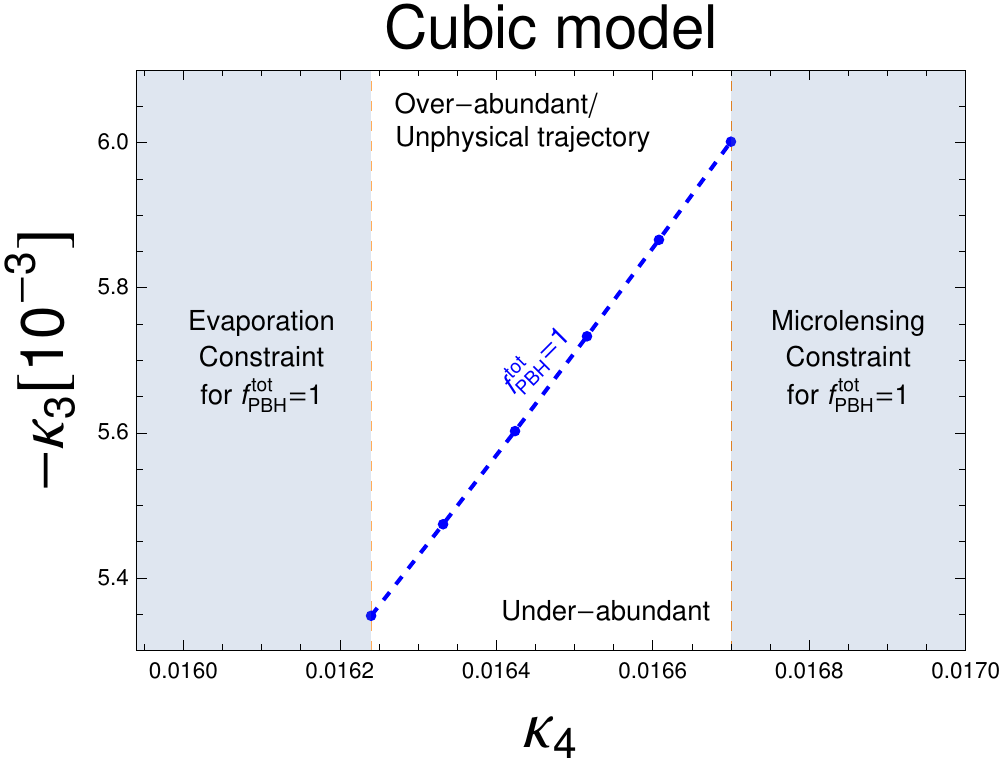}%
\includegraphics[scale=0.85]{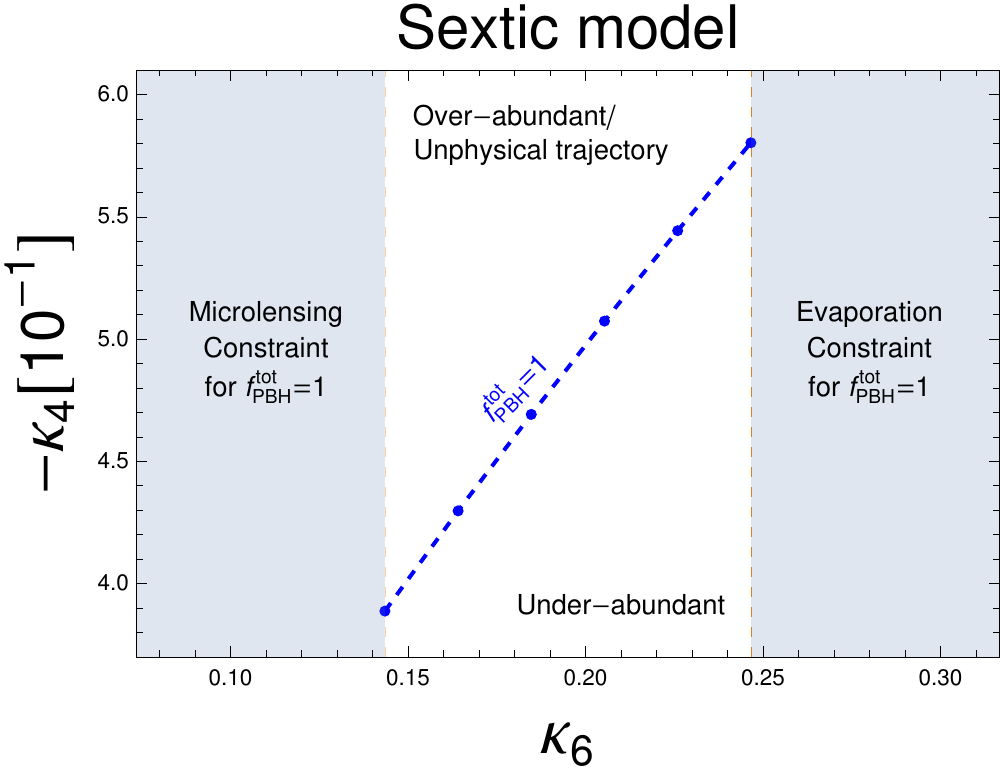}%
\caption{\label{fig:kappas}
The dashed line indicates the parameters $(\kappa_4,\kappa_3)$ of the cubic mode (the left panel) and the $(\kappa_6,\kappa_4)$ of the sextic model giving $f^{\rm tot}_{\rm PBH}=1$, namely the primordial black holes comprise the total dark matter abundance today. The region below the line corresponds to under-abundance. 
The region above the line corresponds to over-abundant PBHs and the trajectories that fall into one of the non-Standard Model vacua (we call them false vacua) as shown in the rightmost panels of Fig.~\ref{fig:cubictraj} and Fig.~\ref{fig:sextictraj}, but the over-abundance region is so narrow that it is not recognizable.
}
\end{figure*}

Figure~\ref{fig:kappas} shows the dependence on the parameters $\kappa_4$ and $\kappa_3$ for the cubic model case (the left panel) and on the parameters $\kappa_6$ and $\kappa_4$ for the sextic model case (the right panel).
With these parameters the peak frequency of the enhanced perturbations shifts, and thus the resulting PBH mass spectrum also changes. The observationally viable parameters are then constrained by the evaporation bounds and the microlensing bounds, see Fig.~\ref{fig:pbh}.
We have found $0.01624 \lesssim \kappa_4 \lesssim 0.01670$ for the cubic model and $0.1435 \lesssim \kappa_6 \lesssim 0.2466$ for the sextic model.
In both panels, the dashed line indicates the model parameters yielding the PBHs corresponding to the total dark matter abundance ($f^{\rm tot}_{\rm PBH}=1$).
The region under the line corresponds to $f^{\rm tot}_{\rm PBH}<1$.
The region above the line corresponds to trajectories that fall into one of the minima that does not represent our Universe.
There is a region right above the line that gives over-abundant PBH $f^{\rm tot}_{\rm PBH}>1$, which is too narrow to be visible.

As one can see from Fig.~\ref{fig:kappas}, the production rate $f^{\rm tot}_{\rm PBH}$ is rather sensitive to the choice of the K\"{a}hler potential parameters $\kappa_3$, $\kappa_4$, $\kappa_6$. 
This type of sensitivity is a common feature in similar inflationary production mechanisms of primordial black holes.
Since our scenario is based on supersymmetry, the prediction, while sensitive to the parameters, is robust against radiative corrections compared to other scenarios without supersymmetry.

\subsection{Induced Gravitational Waves}
In the subhorizon region the gravitational waves have the energy density of \cite{Maggiore:1999vm,Maggiore:2007ulw}
\begin{align}\label{eqn:GWdensity}
\rho_{\rm GW} = \frac{1}{16a^2}\langle
\overline{
\partial_k h_{ij}
\partial^k h^{ij}}
\rangle\,,
\end{align}
where the overline denotes the average over oscillations.
Let us decompose the tensor $h_{ij}$ as
\begin{align}\label{eqn:FTh}
h_{ij}(t,{\bf x}) = \int \frac{d^3k}{(2\pi)^{3/2}}\left(
h_{\bf k}^+(t)e_{ij}^+({\bf k}) + h_{\bf k}^\times(t)e_{ij}^\times({\bf k})
\right)e^{i{\bf k}\cdot{\bf x}}\,,
\end{align}
where
\begin{align}
e_{ij}^+({\bf k}) &= \frac{1}{\sqrt{2}}\left(
e_i({\bf k})e_j({\bf k}) - \bar{e}_i({\bf k})\bar{e}_j({\bf k})
\right) \,,\\
e_{ij}^\times({\bf k}) &= \frac{1}{\sqrt{2}}\left(
e_i({\bf k})\bar{e}_j({\bf k}) + \bar{e}_i({\bf k})e_j({\bf k})
\right)\,,
\end{align}
are the polarization tensors with $e_i({\bf k})$ and $\bar{e}_i({\bf k})$ being two orthogonal unit vectors. Using the expression \eqref{eqn:FTh} in the energy density \eqref{eqn:GWdensity} gives
\begin{align}
\rho_{GW}(t) = \int d\ln k \, \frac{1}{8}\left(\frac{k}{a}\right)^2
\overline{\mathcal{P}_h(t,k)}\,.
\end{align}
Here $\mathcal{P}_h(t,k) \equiv \mathcal{P}^{+,\times}_h(t,k)$, defined by
\begin{align}
\langle h_{\bf k}^+(t)h_{\bf q}^+(t) \rangle &=
\delta^3({\bf k} + {\bf q})\frac{2\pi^2}{k^3}\mathcal{P}^+_h(t,k)\,,\\
\langle h_{\bf k}^\times(t)h_{\bf q}^\times(t) \rangle &=
\delta^3({\bf k} + {\bf q})\frac{2\pi^2}{k^3}\mathcal{P}^\times_h(t,k)\,.
\end{align}
We note that $\mathcal{P}^+_h(t,k) = \mathcal{P}^\times_h(t,k)$.
Omitting the polarization index, we define the gravitational wave energy density parameter as follows:
\begin{align}
\Omega_{\rm GW}(t,k) \equiv \frac{\rho_{\rm GW}(t,k)}{\rho_{\rm crit}}
= \frac{1}{24}\left(\frac{k}{aH}\right)^2\overline{\mathcal{P}_h(t,k)}\,.
\end{align}

\begin{figure*}[ht!]
\centering
\includegraphics[width=0.495\textwidth]{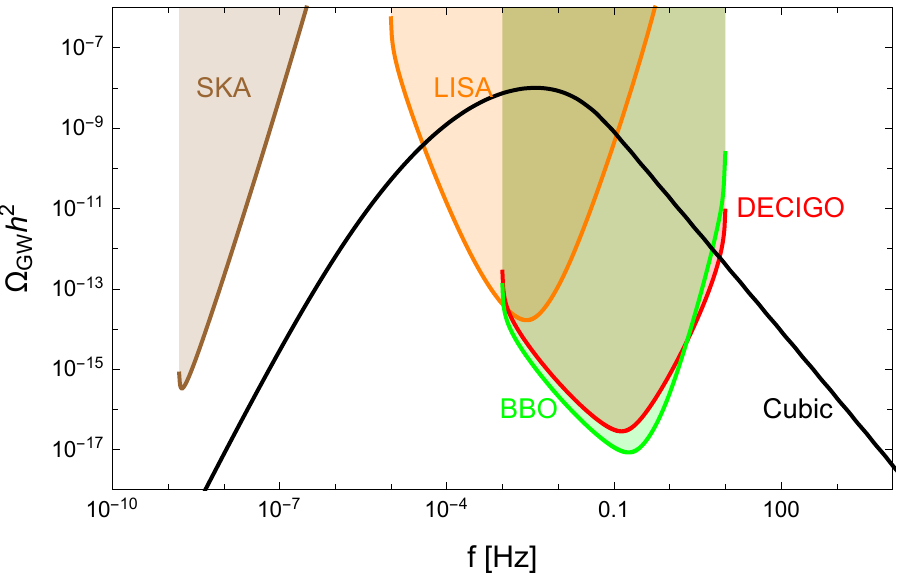}
\includegraphics[width=0.495\textwidth]{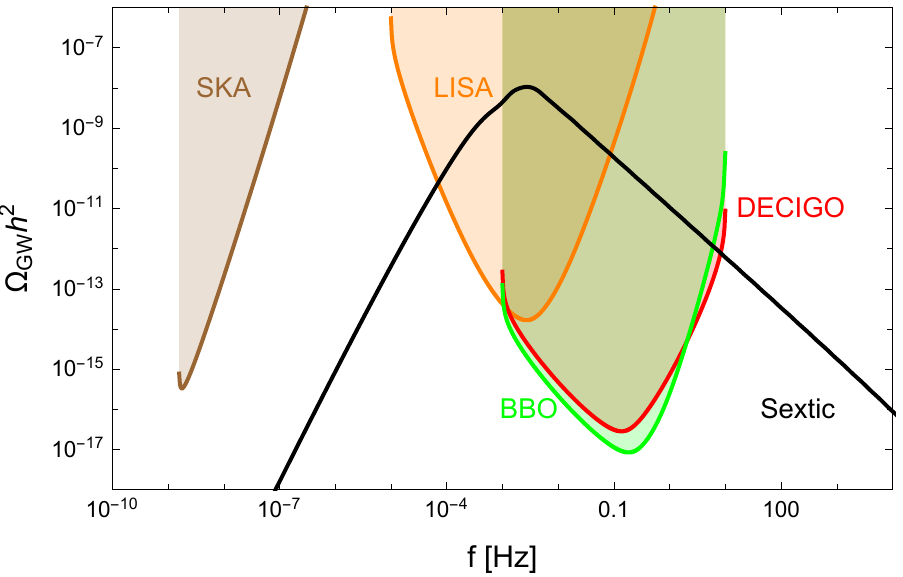}
\caption{The spectrum of the gravitational waves generated from the curvature perturbations. The left panel shows the cubic model and the right panel shows the sextic model. The chosen parameters are the same as those presented in Fig.~\ref{fig:pbh}. The data for the sensitivity curves are obtained from \cite{Schmitz:2020syl,Schmitz:2020aaa}.}
\label{fig:gw}
\end{figure*}

To obtain the tensor power spectrum one may solve the tensor perturbation equation \cite{Baumann:2007zm,Kohri:2018awv,Inomata:2016rbd}
\begin{align}
h_{\bf k}^{\prime\prime} + 2aHh_{\bf k}^\prime + k^2h_{\bf k} = 4S_{\bf k}\,,
\end{align}
where the prime denotes the derivative with respect to the conformal time. The source term $S_{\bf k}$ is given by
\begin{align}
S_{\bf k} &= \int \frac{d^3q}{(2\pi)^{3/2}} e_{ij}({\bf k})q_iq_j\bigg[
2\Psi_{\bf q}\Psi_{{\bf k}-{\bf q}}
\\
&\quad
+\frac{4}{3(1+w)}(\mathcal{H}^{-1}\Psi_{\bf q}^\prime + \Psi_{\bf q})
(\mathcal{H}^{-1}\Psi_{{\bf k}-{\bf q}}^\prime + \Psi_{{\bf k}-{\bf q}})
\bigg]\,,\nonumber
\end{align}
where $w$ is the equation of state and $\Psi_{\bf k}$ is the Fourier transform of the scalar perturbation in the conformal Newtonian gauge where the metric is given by $ds^2=-(1+2\Psi)dt^2+a^2[(1-2\Psi)\delta_{ij}+h_{ij}/2]dx^idx^j$ with the anisotropic stress tensor being neglected.
We assume that the induced gravitational waves are generated in the radiation dominated era. Then, at the generation time, we obtain \cite{Kohri:2018awv}
\begin{align}\label{eqn:OmegaGWf}
&\Omega_{\rm GW}(t_{\rm f},k) =
\frac{1}{12}\int_0^\infty dv
\int_{|1-v|}^{1+v} du
\nonumber\\
&\quad\times
\left(
\frac{4v^2 - (1+v^2-u^2)^2}{4uv}
\right)^2
\nonumber\\
&\quad\times
\mathcal{P}_\zeta(kv)\mathcal{P}_\zeta(ku)
\left(
\frac{3(u^2+v^2-3)}{4u^3v^3}
\right)^2
\nonumber\\
&\quad\times
\bigg[
\left(
-4uv + (u^2+v^2-3)\log\bigg\vert
\frac{3-(u+v)^2}{3-(u-v)^2}
\bigg\vert
\right)^2 
\nonumber\\
&\hspace{2cm}
+ \pi^2(u^2+v^2-3)^2\theta(v+u-\sqrt{3})
\bigg]\,.
\end{align}
Multiplying the present day radiation energy density parameter $\Omega_{{\rm rad},0}$, we find the energy density parameter today $\Omega_{\rm GW} = \Omega_{\rm rad,0}\Omega_{\rm GW}(t_{\rm f})$ \cite{Kohri:2018awv,Ando:2018qdb}.

The spectrum of the scalar-sourced second order gravitational waves generated from the cubic and sextic models is shown in Fig.~\ref{fig:gw} together with the sensitivity curves for future experiments such as LISA \cite{LISA:2017pwj,Baker:2019nia}, DECIGO \cite{Seto:2001qf,Kawamura:2006up,Sato:2017dkf,Isoyama:2018rjb,Kawamura:2020pcg}, BBO \cite{Corbin:2005ny,Crowder:2005nr,Harry:2006fi}, and SKA \cite{Carilli:2004nx,Janssen:2014dka,Weltman:2018zrl}. We observe that the signals are within the reach of LISA, DECIGO, and BBO.

\section{Final remarks}
\label{sec:conc}
In this work we analyzed models of two-field inflation which are well motivated from high energy physics and particle phenomenology. 
We demonstrated by numerical studies that these models can produce primordial black holes during the radiation dominated era which are of interest as a candidate of the present day dark matter.
We investigated two sample cases in detail, one with the K\"{a}hler potential including the cubic term of the singlet field, and the other with the sextic term of the singlet field.
We have found that in both cubic and sextic cases significant enhancement of the curvature perturbation can be achieved. 
In both the cubic and sextic cases the enhancement can be large enough so that the produced primordial black holes are abundant enough to account for the whole present day dark matter abundance.
We have also computed the spectrum of the gravitational waves resulting as the secondary effect of the large scalar perturbation.
We have found that, for both cubic and the sextic cases, the model can generate gravitational waves in the target range of detectors, including LISA, DECIGO, and BBO.

The focus of this paper has been on the features of the K\"{a}hler potential and the trajectory of the inflaton that may lead to the production of primordial black holes.
While we used the simple superpotential \eqref{eqn:W}, this may be generalized to more realistic phenomenological examples.
It would be certainly interesting to analyze various cases based on specific particle physics models.
For example, in the $SU(5)$ GUT model \cite{Kawai:2015ryj} the inflaton trajectory needs to settle to the Standard Model vacuum with broken $SU(5)$ symmetry, and it would be interesting to investigate whether such trajectories are compatible with the primordial black hole production. 
We plan to examine such questions in the future.

\begin{acknowledgments}
J. K. would like to thank Dhiraj Kumar Hazra and Sarah Geller for useful discussions on multifield inflation.
This work was supported in part by the National Research Foundation of Korea Grant-in-Aid for Scientific Research NRF-2022R1F1A1076172 (SK).
\end{acknowledgments}



\end{document}